%% file: main.tex
\documentclass[conference,9pt]{IEEEtran}
\input{misc/customize}

\advance\topmargin-1in\advance\oddsidemargin-1in
\evensidemargin\oddsidemargin

\makeatletter
\def\@normalsize{\@setsize\normalsize{12pt}\xpt\@xpt
\abovedisplayskip 10pt plus2pt minus5pt\belowdisplayskip \abovedisplayskip
\abovedisplayshortskip \z@ plus3pt\belowdisplayshortskip 6pt plus3pt
minus3pt\let\@listi\@listI}

\def\section{\@startsection {section}{1}{\z@}{20pt plus 2pt minus 2pt}
{8pt plus 2pt minus 2pt}{\centering\normalsize\sc
\edef\@svsec{\thesection.\ }}}
\def\thesection{\Roman{section}}

\def\subsection{\@startsection {subsection}{2}{\z@}{16pt plus 2pt minus 2pt}
{6pt plus 2pt minus 2pt}{\normalsize\sl
\edef\@svsec{\thesubsection.\ }}}
\def\thesubsection{\Alph{subsection}}

\long\def\@makecaption#1#2{
\vskip10pt\begin{center} #1 #2 \end{center}\par\vskip 1pt}
\def\fnum@figure{\raggedright{\footnotesize Fig. \thefigure }.%
\footnotesize}
\def\fnum@table{\footnotesize TABLE \thetable\\\footnotesize\sc}
\def\thetable{\Roman{table}}

\makeatother

\begin{document}
\date{}

\title{DROID: Discrete-Time Simulation 
for Ring-Oscillator-Based Ising Design}

\author{Abhimanyu Kumar\textsuperscript{1}, Ramprasath S.\textsuperscript{2}, Chris H. Kim\textsuperscript{1}, Ulya R. Karpuzcu\textsuperscript{1}, Sachin S. Sapatnekar\textsuperscript{1}\\
\small{\textsuperscript{1} University of Minnesota, Minneapolis, USA \textsuperscript{2} Indian Institute of Technology Madras, Chennai, India}}

\maketitle

\begin{abstract}
Many combinatorial problems can be mapped to Ising machines, i.e., networks of coupled oscillators that settle to a minimum-energy ground state, from which the problem solution is inferred. This work proposes DROID, a novel event-driven method for simulating the evolution of a CMOS Ising machine to its ground state. The approach is accurate under general delay-phase relations that include the effects of the transistor nonlinearities and is computationally efficient. On a realistic-size all-to-all coupled ring oscillator array, DROID is nearly four orders of magnitude faster than a traditional HSPICE simulation in predicting the evolution of a coupled oscillator system and is demonstrated to attain a similar distribution of solutions as the hardware.
\end{abstract}

\input{sec/1_Introduction}
\input{sec/2_Background}
\input{sec/3_Relation_to_genAdler}

\input{sec/4_Simulation}
\input{sec/5_Results}
\input{sec/6_Conclusion}

\bibliographystyle{misc/IEEEtran}

\appendices
\renewcommand{\thesubsection}{\Roman{subsection}}
\input{sec/Appendix}

\end{document}

%% file: misc/customize.tex
\usepackage{graphicx}
\usepackage{multirow}
\usepackage{multicol}
\usepackage{color}
\usepackage{url}
\usepackage{array}
\newcolumntype{P}[1]{>{\centering\arraybackslash}p{#1}}
\usepackage{algorithm}
\usepackage{comment}
\usepackage{algpseudocode}
\usepackage{mathtools} 
\usepackage[export]{adjustbox}
\usepackage{textcomp}
\usepackage{cite}
\usepackage{amsmath}
\usepackage{amssymb}
\usepackage{mwe}

\usepackage{subfigure}
\usepackage[normalem]{ulem}
\usepackage{enumitem}
\usepackage{float}
\usepackage{enumitem}
\definecolor{gray1}{gray}{0.7}
\definecolor{gray2}{gray}{0.98}
\definecolor{light-gray}{gray}{0.95}
\newcommand{\ignore}[1]{}

\newcommand{\phase}[3][]{\phi_{#1#2}^{#3}}
\newcommand{\period}[3][]{T_{#1#2}^{#3}}
\newcommand{\freq}[3][]{\omega_{#1#2}^{#3}}

%% file: sec/1_Introduction.tex
\section{Introduction}
\label{sec:Intro}

\noindent
A critical computational domain for hardware accelerators is the area of solving combinatorial optimization problems (COPs) that are NP-complete or NP-hard -- e.g., the traveling salesman, satisfiability, and knapsack problems.  Today, such problems are solved on classical computers using heuristics with no optimality guarantees, or approximation algorithms with loose optimality bounds.

Ising computation is a promising emerging computational model for solving COPs.  Ising machines are inspired by the work of Ernst Ising, who proposed a formulation based on binary states called {\em spins}, with allowable values of $+1$ and $-1$, to explain ferromagnetism. Such systems have a natural tendency to find a {\em ground state} with a configuration of spins that minimizes energy.  Ising computation maps discrete combinatorial optimization problems to this paradigm.  Under a linear transformation, Boolean quadratic unconstrained binary optimization (QUBO) problems can be formulated as two-body Ising interactions.   Karp's list of 21 NP-complete problems are shown to have an Ising formulation~\cite{Lucas_Ising_Frontiers14}, and many other problems can also be formulated in this way. 

Recently, there has been great interest in building Ising hardware accelerators, realizing spins using superconducting loops in D-Wave machines~\cite{Johnson2011, Bian2014}; optical parametric oscillators in coherent Ising machines~\cite{Inagaki2016,Yamamoto2017}, ring oscillators (ROs) in CMOS-based Ising machines~\cite{wang2019matlab,moy20221,Lo2023},
and memory cells in SRAM-based engines~\cite{Yamaoka16}.   
Oscillator-based methods use the phenomenon of {\em synchronization}, whereby a system of coupled oscillators with similar frequencies, converge to a common frequency and fixed phase difference through injection locking.
The dynamics of coupled oscillators have been studied as early as 1663, when Huygens noticed the synchronization of pendulums connected to a common bar~\cite{Willms17}. Adler derived a closed-form expression for locking in LC oscillators~\cite{Adler}, while Winfree explored weak interactions of periodic behavior in biological rhythms~\cite{WINFREE196715}. 

Kuramoto's analysis~\cite{Kuramoto1984-hj} studied chemical oscillations under sinusoidal interactions. These works simulate synchronization behavior 
through differential equations that relate the rate of change of each oscillator phase to the phases and frequencies of other oscillators. 

Unlike platforms using exotic futuristic technologies, CMOS RO-based Ising machines use a mainstream semiconductor technology that is scalable, compact, economically and reliably mass-manufacturable today, and can operate at room temperature instead of requiring expensive high-power mK-level refrigeration schemes. The synchronization of RO-based Ising machines can be simulated using HSPICE, but this is computationally intensive and does not scale well. Simulators for oscillator-based Ising machines are based on analytical solutions to the generalized Adler equation~\cite{Bhansali2009} and the generalized Kuramoto equation~\cite{wang2019matlab}.  A prior event-driven approach~\cite{sreedhara23} fast-forwards through multiple RO cycles until the phase difference between some pair of ROs crosses an integer multiple of $\pi$: this is considered to be an event. If the phase difference remains close to an integer multiple of $\pi$ for some iterations, the associated coupling is removed from the system and a phase merging scheme is used to lock the phases of these oscillators henceforth. A hardware realization of a generalized Kuramoto equation solver has also been demonstrated~\cite{sreedhara_date23}.

Prior methods have several limitations.  First, 
they represent the phase of each oscillator by the phase of a single reference stage. However, the phase differences at specific coupling sites between two oscillators may differ from the differences in their reference phases. 
Second, methods that use phase merging~\cite{sreedhara23} can be misleading: the phase of an RO can diverge even after it appears to come close to another RO phase. 
This work proposes DROID (Discrete-Time Simulation for Ring-Oscillator-Based Ising Design), a method for simulating RO-based Ising machines, that overcomes the above limitations. Its contributions are as follows:
\begin{itemize}[noitemsep,topsep=-1pt,leftmargin=*]
\item 
We show that for coupled RO systems, prior continuous-time (CT) simulation abstractions, such as the generalized Adler formulation~\cite{Bhansali2009}, are abstractions of a discrete-event simulation, operating under restrictive assumptions that allow closed-form solutions, including assumptions of infinitesimal changes 
(Section~\ref{sec:Adler_relation}).  Our approach removes these restrictions and uses lookup-table-based functions, characterized using HSPICE. 

\item 
Unlike prior methods that work in the continuous domain, we develop a discrete-time event-driven simulation methodology (Section~\ref{sec:Simulation}) to predict the behavior of coupled RO systems; this method is inspired by timing analysis methods that are widely used for digital circuits, which achieve acceptable accuracy at a fraction of the runtime of HSPICE. Our approach is event-driven, where an event is defined with fine granularity, associated with a coupled transition between two oscillators.

\item 
Our approach is 125$\times$--7441$\times$ faster than HSPICE at similar accuracy, with larger speedups for larger systems. We match the distribution of our solutions, across 250 problems of various oscillator coupling densities, 100 samples per problem, and multiple initial conditions, against a CMOS RO-based Ising hardware solver~\cite{Lo2023}, and show that the distance between distributions, is small. 

\end{itemize}

The paper is organized as follows. Section~\ref{sec:Background} summarizes the concepts that guide this work. Section~\ref{sec:A2A} describes an all-to-all-connected Ising hardware accelerator that serves as our hardware testcase. Sections~\ref{sec:Adler_relation} and~\ref{sec:limitations_of_ct_approx} then analyze the relationship between discrete-time and continuous-time simulation of coupled oscillator systems.  We describe our event-driven simulation scheme for the Ising hardware in Section~\ref{sec:Simulation} and show our simulator results in Section~\ref{sec:Results}, finally concluding the paper in Section~\ref{sec:Conclusion}.

%% file: sec/2_Background.tex
\section{CMOS-based Coupled-oscillator Systems}
\label{sec:Background}

\noindent
\subsection{The Ising model}
\label{subsec:Ising}

\noindent
The Ising formulation of a COP minimizes the following objective function, referred to as a {\em Hamiltonian}:
\begin{equation}
    H(\mathbf{s}) = \textstyle -\sum_{i = 1}^{N} \sum_{j=1}^{N} J_{ij} s_i s_j - \sum_{i=1}^{N} h_i s_i,
    \label{eq:ham}
\end{equation}
In the magnetics domain, this models the energy of a system of $N$ spins; spin is an intrinsic property associated with a subatomic particle, atom, or molecule, and can take on a value of $+1$ or $-1$.
The Hamiltonian is the energy of a system of spins as a function of their interactions ($J_{ij} s_i s_j$) and the effect of external magnetic fields on individual spins ($h_i s_i$). A physical Ising machine settles to a ground state of low-energy states favored by nature, thus minimizing the Hamiltonian. Therefore, by suitably mapping a COP to the weights $J_{ij}$ and $h_i$, an Ising machine can solve a COP formulated as a Hamiltonian.

Two spins $s_i$ and $s_j$ are in-phase if $s_i = s_j$, and out-of-phase otherwise. From~\eqref{eq:ham}, a positive [negative] $J_{ij}$ encourages $s_i$ and $s_j$ to be in-phase [out-of-phase], and a positive [negative] $h_i$ pushes $s_i$ to be $+1$ [$-1$].

\subsection{CMOS-Ring-Oscillator-Based Ising Machines}
\label{subsec:ro_ising}

\vspace{-4mm}
\begin{figure}[htb]
    \centering
    \includegraphics[width=0.8\linewidth]{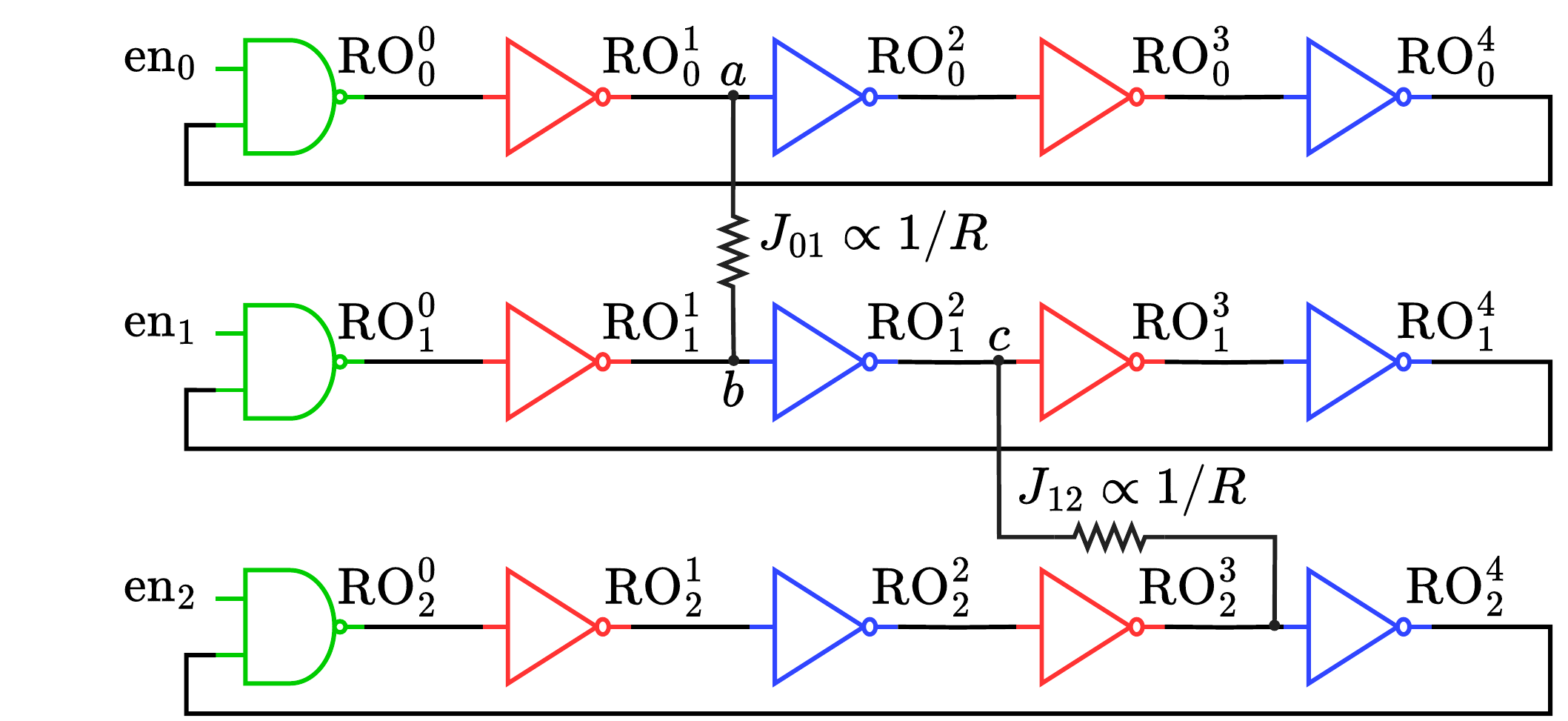}
    \vspace{-4mm}
    \caption{Three five-stage ROs, with a positive and a negative coupling.  The green stage is the reference, and odd (even) stages are shown in red (blue).}
    \label{fig:RO_coupling}
    \vspace{-2mm}
\end{figure}

\noindent
\underline{Core principle.}
Fig.~\ref{fig:RO_coupling} shows a CMOS-based Ising machine with three coupled ROs. Each RO has an identical number of stages; each stage in the $i^{\rm th}$ oscillator, RO$_i$, is an inverter, except for one NAND stage with an enable signal, en$_i$, to start the oscillator. 

We denote the $k^{\rm th}$ stage of RO$_i$ as RO$_{i}^{k}$, with a phase, $\phi_i^k$, given by the arrival time of the rising edge at the stage output. The phase $\phi_i$ of an oscillator is defined as the phase, $\phi_i^0$, at the output of its zeroth stage (shown in green in the figure). The phase at stage RO$_i^k$ is found by adding to $\phi_i$ the sum of the stage delays from RO$_i^0$ to RO$_i^k$.  The time between two consecutive rising edges at the output of a stage is the period of the RO. We denote the nominal period, i.e., the period of each uncoupled, freely oscillating RO, as $T$. When the oscillators are coupled together, they synchronize to the same period, which may be different from $T$.

We designate one of the oscillators as a reference oscillator, always setting its spin to $+1$; without loss of generality, we refer to this as RO$_0$. To assign a spin value to an oscillator, its phase is compared with that of the reference oscillator: an oscillator that is in-phase with RO$_0$ is said to have a spin of $+1$, and one that is out-of-phase has a spin of $-1$.

Ising machines employ weak coupling~\cite{Kuramoto1984-hj}, where the delay change due to coupling is small compared to the nominal period $T$. A non-zero coupling coefficient, $J_{ij}$, in the Ising model is realized by coupling RO$_i$ and RO$_j$. One way to couple a pair of ROs is by connecting the outputs of two corresponding stages in each RO by a resistor; other coupling schemes may also be used~\cite{cilasun2024}. For resistive coupling, the coupling strength $J_{ij}$ is determined by the values of the resistors between inverters in RO$_i$ and RO$_j$; Section~\ref{sec:A2A} will describe a circuit that implements multiple programmable $J_{ij}$ values.  

We refer to inverters that are at an even-parity [odd-parity] distance from the reference stage as even [odd] stages, shown in blue [red] in Fig.~\ref{fig:RO_coupling}. Coupling between two same-parity stages in different ROs is referred to as {\em positive coupling} (RO$_0$ and RO$_1$ in Fig.~\ref{fig:RO_coupling}), while coupling between opposite-parity stages is termed {\em negative coupling} (RO$_1$ and RO$_2$ in Fig.~\ref{fig:RO_coupling}). Positive coupling encourages the ROs to be in-phase, while negative coupling encourages the ROs to be out-of-phase.
In Fig.~\ref{fig:RO_coupling}, the net $a$ driven by stage RO$_0^1$ is coupled to net $b$ at the output of RO$_1^1$ by a resistor; this causes the stage delays to change from their nominal values:
\begin{itemize}[noitemsep,topsep=-1pt,leftmargin=*]
\item 
When $a$ rises, if $b$ is low (i.e., yet to rise), it opposes the rise transition and the delay of $a$ is increased by $\delta_1^a$ relative to the uncoupled case.  On the other hand, if $b$ is high (i.e., it has already risen), it aids the rise and reduces the delay of $a$ by $\delta_2^a$.  In both cases, the rising edge of $a$ is brought closer to the rising edge of $b$, reducing the phase difference between the signals. 
\item
Similarly, when $a$ falls, if $b$ is high (i.e., yet to fall), the delay is increased by $\delta_3^a$; if $b$ is low (i.e., has already fallen), its delay reduces by $\delta_4^a$. In each case, the falling edges of $a$ and $b$ are brought closer, bringing the oscillators closer to phase-locking.
\end{itemize} 
Net $b$ behaves analogously, with delay shifts of $\delta_1^b, \cdots , \delta_4^b$.   

\begin{figure}[htb]
\centering
\subfigure[]{\includegraphics[width=0.45\textwidth]{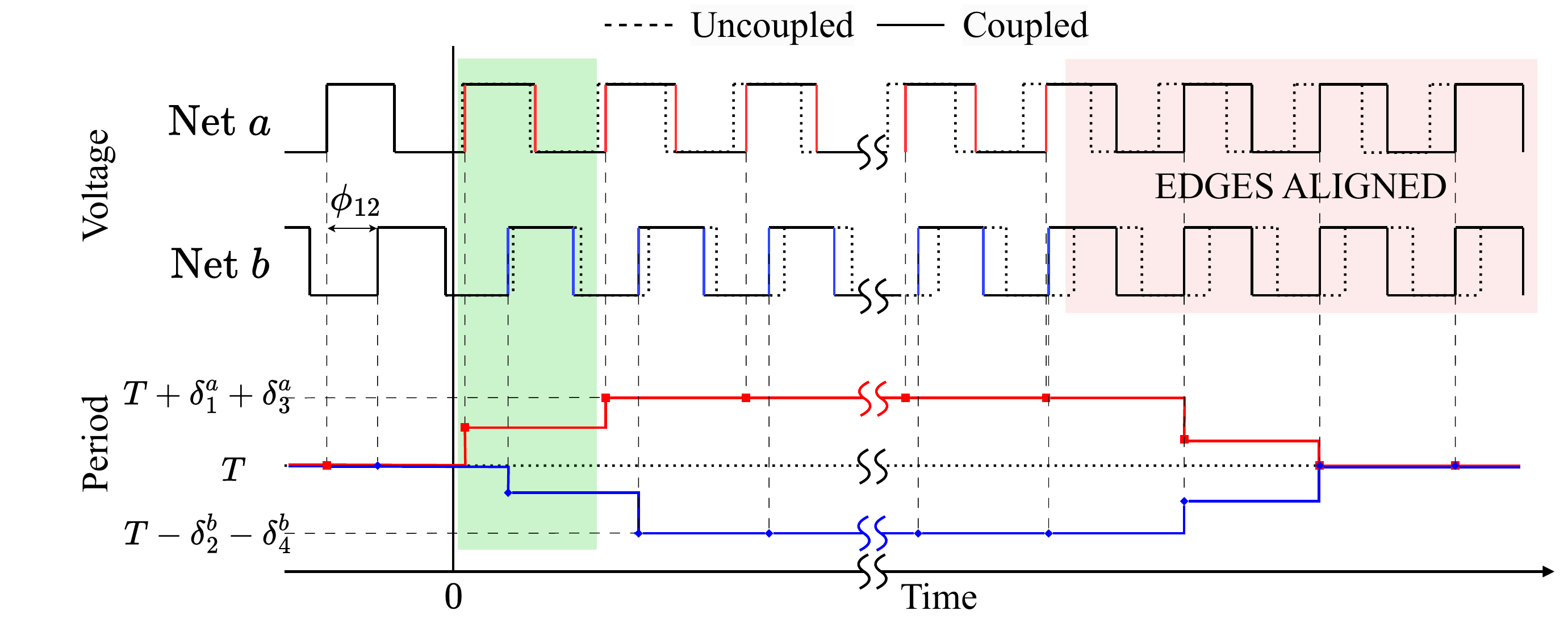}
\label{fig:coupled_waveform}}

\vspace{-4mm}
\subfigure[]{\includegraphics[width=0.35\textwidth]{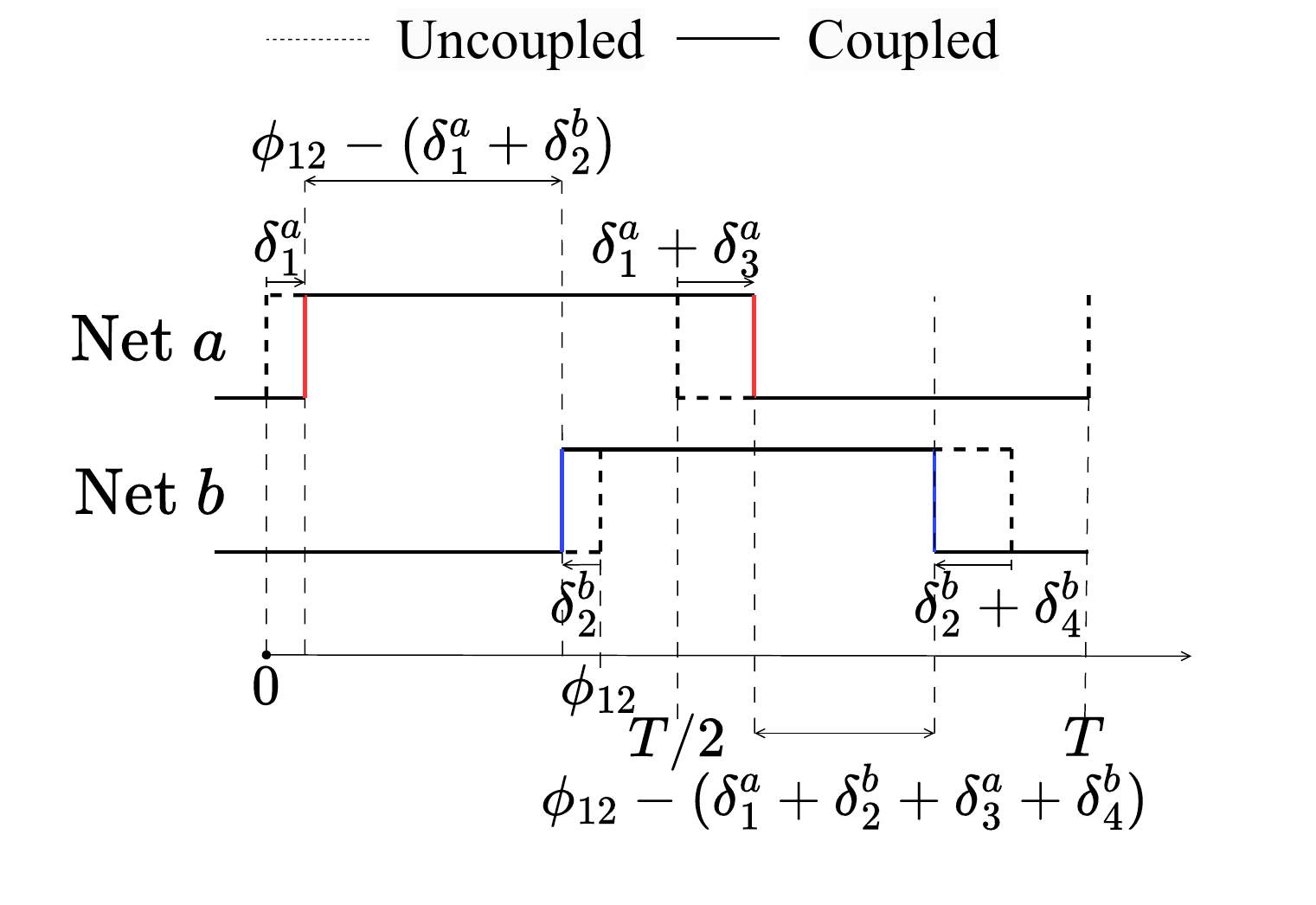}
\label{fig:effect_of_coupling}}
\vspace{-3mm}
\caption{(a)~Waveforms for nets $a$ and $b$ in two ROs, when uncoupled, and with coupling enabled at $t=0$. (b)~Detailed view of the green box in (a), showing the effect of coupling. The period of waveform at $a$ increases while that at $b$ decreases, reducing the phase difference.}
\label{fig_sim}
\vspace{-6mm}
\end{figure}

Consider the coupled oscillator system of Fig.~\ref{fig:RO_coupling}, but with the coupling between RO$_1$ and RO$_2$ removed, i.e., only RO$_0^1$ and RO$_1^1$ are coupled.  Fig.~\ref{fig:coupled_waveform} shows the effect of coupling over multiple cycles, where the waveforms at $a$ and $b$ begin with a phase difference, $\phi_{12}$. In the uncoupled case, this phase difference is unchanged, but under coupling, as $a$ slows down and $b$ speeds up, the phase difference decreases and the edges align. The inset in Fig.~\ref{fig:effect_of_coupling} shows the first cycle, i.e., the green region in Fig.~\ref{fig:coupled_waveform}. Due to coupling, the first rising edge of $a$ is delayed by $\delta_1^a$ while that of $b$ arrives earlier by $\delta_2^b$.  Similarly, the falling edge at $a$ is delayed by an additional $\delta_3^a$ while that at $b$ is sped up by an additional $\delta_4^b$. Thus, the phase difference is reduced by $(\delta_1^a + \delta_2^b + \delta_3^a + \delta_4^b)$ after the first cycle. Subsequently, as long as transitions in $a$ are completed before those in $b$, $a$ continues to be delayed by $\delta_1^a+\delta_3^a$ and $b$ continues to speed up by $\delta_2^b+\delta_4^b$ every cycle, and the phase difference at the end of $k$ cycles reduces to $\phi_{12} - k(\delta_1^a + \delta_2^b + \delta_3^a + \delta_4^b)$ until the phases align.

\noindent
\underline{Coupled oscillator systems.}
We abstract a general coupled oscillator system with a graph $G=(V,E)$, where the vertex set $V$ is the set of oscillators, and each element $e_{ij}$ of the edge set $E$ corresponds to a pair of oscillators RO$_i$ and RO$_j$ with an edge weight corresponding to the coupling strength $J_{ij}$. We denote the change in delay of the coupled stage of RO$_i$ in each cycle, under a phase difference of $\phi_{ij}$, by $f_{J_{ij}}(\phi_{ij})$; for the example in Fig.~\ref{fig:effect_of_coupling}, $f_{J_{12}}(\phi_{12}) = (\delta_1^a + \delta_3^a)$ and $f_{J_{12}}(\phi_{21}) = (\delta_2^b + \delta_4^b)$. The net change in the period of RO$_i$ in the $k^{\rm th}$ cycle, $\mathcal{D}_i^k$, is the sum of changes in the delay of each stage, i.e., $\mathcal{D}_i^k = \textstyle \sum_{(i,j) \in E} f_{J_{ij}}(\phase{ij}{k})$.

\noindent
\underline{Synchronization.} A pair of coupled oscillators that have the same period will have a constant phase difference. Conversely, when all phase differences are constant, it implies that all pairs of coupled oscillators have the same period, a phenomenon referred to as \emph{synchronization}. In Fig.~\ref{fig:coupled_waveform}, since the coupling is purely positive, all phases align upon synchronization. As a result, all stage delays go back to their nominal uncoupled values and the period reverts to the nominal period, $T$.

\subsection{Practical considerations for CMOS-based coupled oscillator systems}
\label{sec:practical_considerations}

\noindent
\underline{Delay change as a function of coupling location}
At a given coupling location between two ROs, the delays from the reference stage of each RO to their coupling stages may be different. As a result, the phase difference at the coupling site may be different from the phase difference of the reference ROs. For example, in Fig.~\ref{fig:RO_coupling}, the path from the reference stage to the coupling site for $J_{12}$ involves two inverter delays in RO$_1$, but three in RO$_2$.  Therefore, the phase difference at this coupling site is not the same as $\phi_{12}$, the phase difference between their reference stages. Prior simulators have not considered this issue. 

Moreover, the stage delay depends on the presence or absence of coupling: ROs with more couplings may have different delays to a coupling site than ROs with fewer couplings. Thus, as the phase changes along an RO, delay shifts from earlier stages affect the phase difference at later stages, and using an identical $\phi_{i}$ at all coupling stages leads to inaccuracies. Many events within a cycle interact subtly to produce a total delay shift for each RO. Such interactions within a cycle are considered in our framework.

\begin{figure}[htb]
    \centering
    \includegraphics[width=0.95\linewidth]{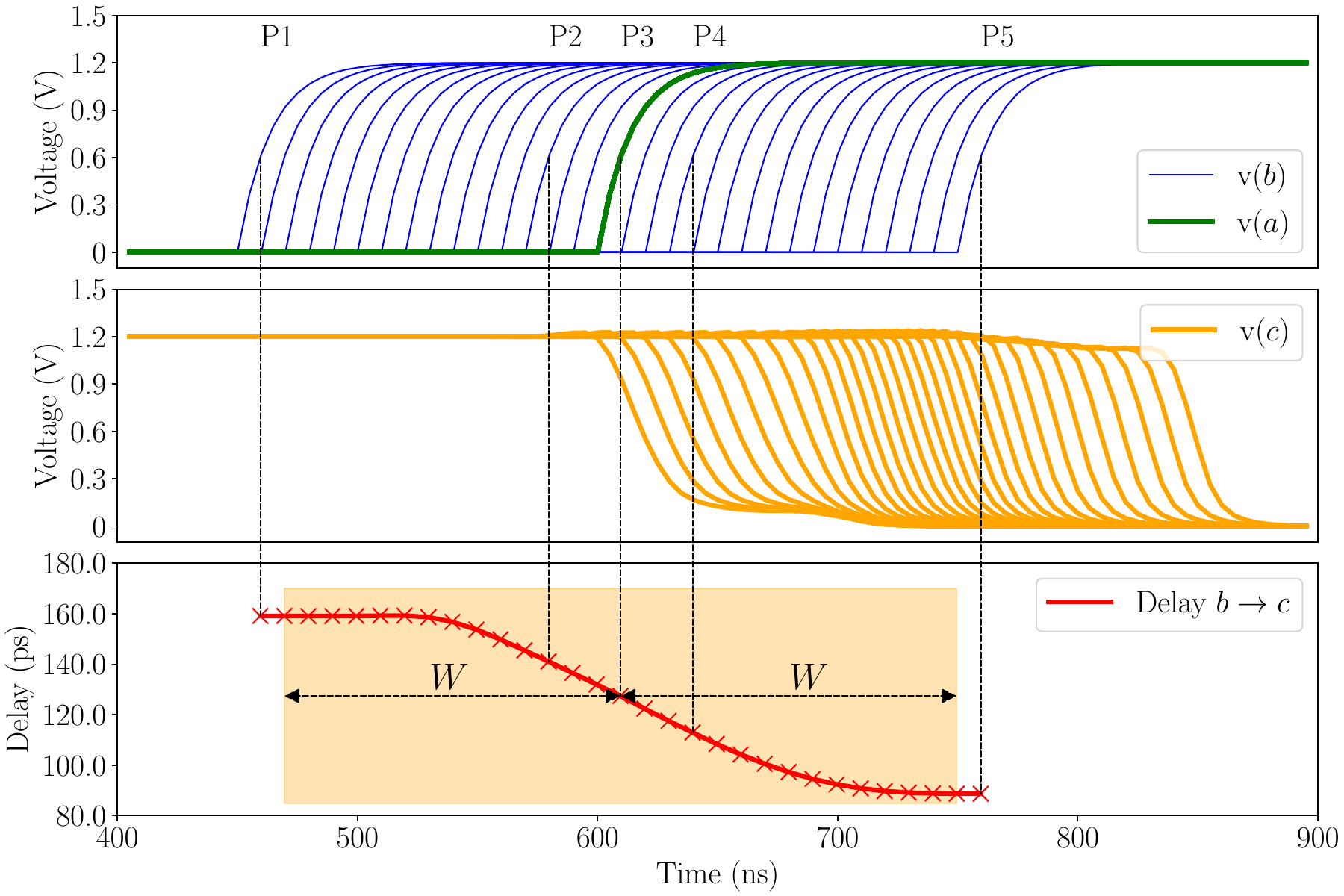} 
    \caption{The relative arrival times on nets $a$ (green waveform) and $b$ (blue waveform), top, impact the transition delay on net $c$ (middle). The transition delay as a function of phase difference, and the interaction window $W$ are illustrated at the bottom.}
    \label{fig:transitions}
\end{figure}

\noindent
\underline{Delay change as a function of arrival time difference}
The speedup or slowdown in stage delay due to coupling depends on the arrival time differences between the signals on the coupled nets. Fig.~\ref{fig:transitions} illustrates this trend at five different relative signal arrival times, (P1, $\cdots$, P5), on nets $a$ (blue) and $b$ (green) in Fig.~\ref{fig:RO_coupling}. Near P1 and P5, the difference in arrival times is large enough that the transition on $a$ does not overlap with a transition on $b$, and the change in delays is constant as $a$ is effectively stable throughout the rise of $b$. At P2 and P4, when the edges are closer, the opposing or assisting 
transistor in the other RO is no longer completely on and it sees a reduced gate-source voltage, which reduces its effect on the delay. If two rising edges with identical transition times are exactly aligned, at P3, no current flows through the resistor, and the transitions do not affect each other. Thus, we see that there is a window around each edge beyond which the arrival of another edge causes a constant stage delay shift, but within this window, the delay shift is a function of the phase difference. This window of width $W$ is called the \emph{interaction window} and extends on either side of a transition as seen by the highlighted orange box in Fig.~\ref{fig:transitions}.

The delay at the output of a stage is also a function of the transition time of the signal at its input~\cite{Sapatnekar04}.  The transition time at net $b$ is also seen to show a similar trend as shown in the bottom panel of Fig.~\ref{fig:transitions}, which has an effect on the delays of subsequent stages.

\section{A silicon-proven all-to-all coupled-RO Ising machine}
\label{sec:A2A}

\begin{figure}[htb]
    \centering
    \includegraphics[width=0.55\textwidth]{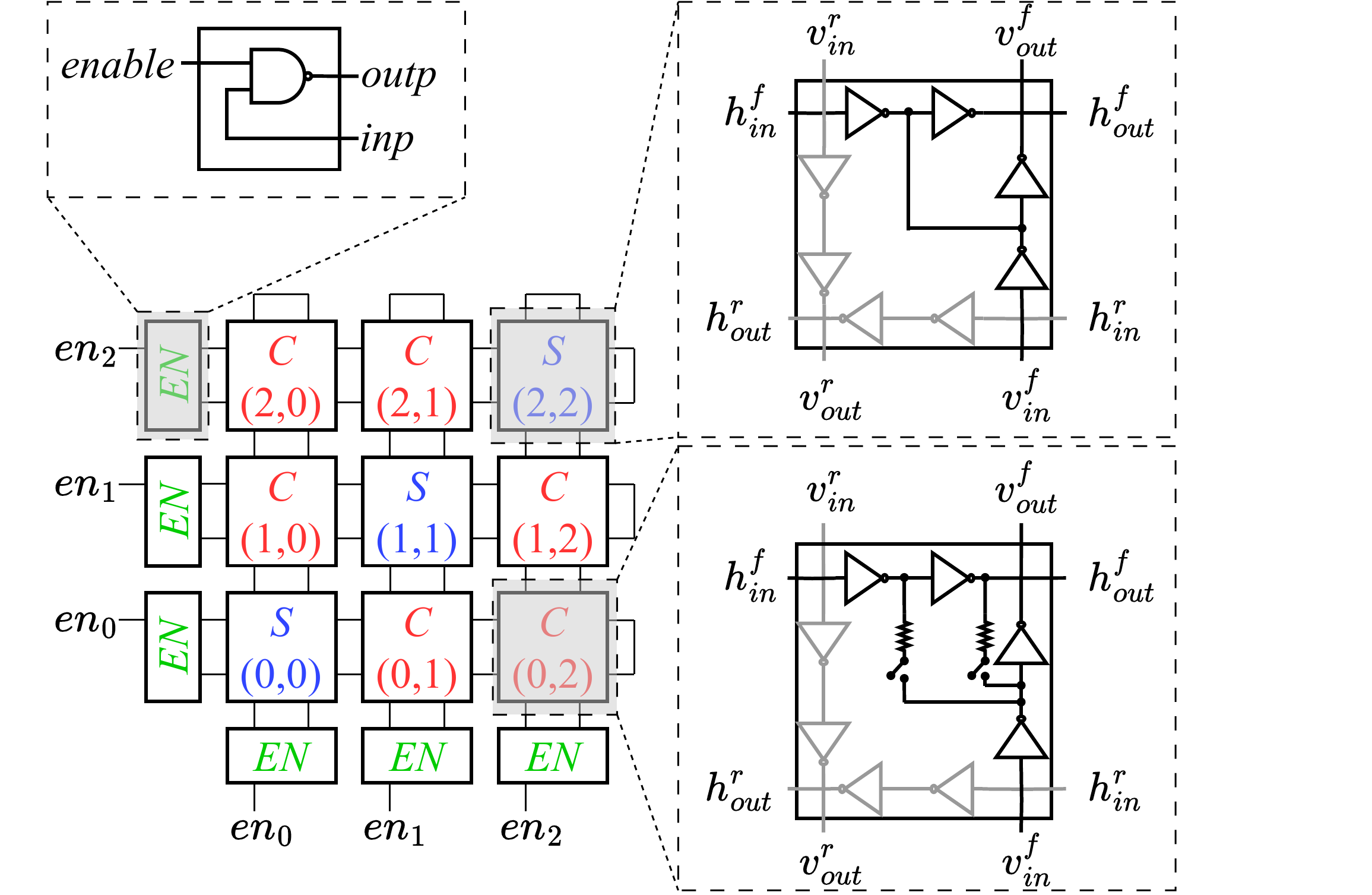}
    \vspace{-6mm}
    \caption{An illustration of the A2A concept through a small three-RO structure, showing shorting cells $S$ on the diagonal, programmable off-diagonal coupling cells $C$, and enable cells $EN$.}
    \label{fig:array_schm}
    \vspace{-4mm}
\end{figure}

\noindent
The design of a coupled-RO Ising machine is particularly tricky because of the need to ensure the uniformity of coupling weights between each pair of ROs. The shift in each transition time in the RO depends on RC parasitics associated with the coupling mechanism, which depend on the precise layout. It has been shown that through regular matched layouts, an all-to-all (A2A) coupled RO array with uniform coupling coefficients can be designed~\cite{Lo2023}.  This A2A design is silicon-proven and will be used in our evaluations.  Note that other designs with planar (hexagonal~\cite{Ahmed2021} and King's graph~\cite{moy20221}) coupling have been proposed, but we focus on the A2A testcase because of its compactness and greater flexibility: in particular, an A2A design with $N$ coupled ROs is equivalent to a planar hexagonal/King's graph array with $\sim N^2$ coupled ROs~\cite{Tabi21, Lucas2019} due to the need for planar arrays to replicate spins during minor embedding~\cite{Lo2023}. This family of A2A arrays has been applied to solve problems ranging from max-cut~\cite{Lo2023} to maximal independent set~\cite{cilasun2024} to satisfiability~\cite{cilasun20243sat}.

For illustration, a simplified schematic, with $J_{ij} \in \{-1,0,+1\}$, for a three-RO system is presented in Fig.~\ref{fig:array_schm}. Each RO is a combination of a vertical and a horizontal RO that are strongly coupled so as to implement the same spin, and have \textit{enable cells} $EN$ (similar to $RO_i^0$ in Fig.~\ref{fig:RO_coupling}) outside the array. Strong couplings are implemented as shorts (upper inset) in the \textit{shorting cells} $S$, placed at the diagonals $(i,i)$. Each \textit{coupling cell} $C$ at off-diagonal location $(i,j)$ has programmable coupling between RO$_i$ and RO$_j$.  A coupling cell has two switches (lower inset) to enable either a positive or a negative coupling. The simplified figure shows possible couplings of $\{-1, 0, +1\}$, but coupling of $\pm 7$ is demonstrated in silicon~\cite{Lo2023}.  Using couplings at both $(i,j)$ and $(j,i)$, each programmable to integer values up to $\pm C_{max}$, the coupling coefficient of $s_i s_j$ can implement integer $J_{ij} \in [-2C_{max}, +2C_{max}]$. 

\vspace{-4mm}

%% file: sec/3_Relation_to_genAdler.tex
\section{Discrete-time vs. continuous-time simulation of coupled-RO systems}
\label{sec:Adler_relation}

\begin{figure}[htb]
    \centering
    \hspace*{-3mm}
    \subfigure[]{\includegraphics[width=0.45\linewidth]{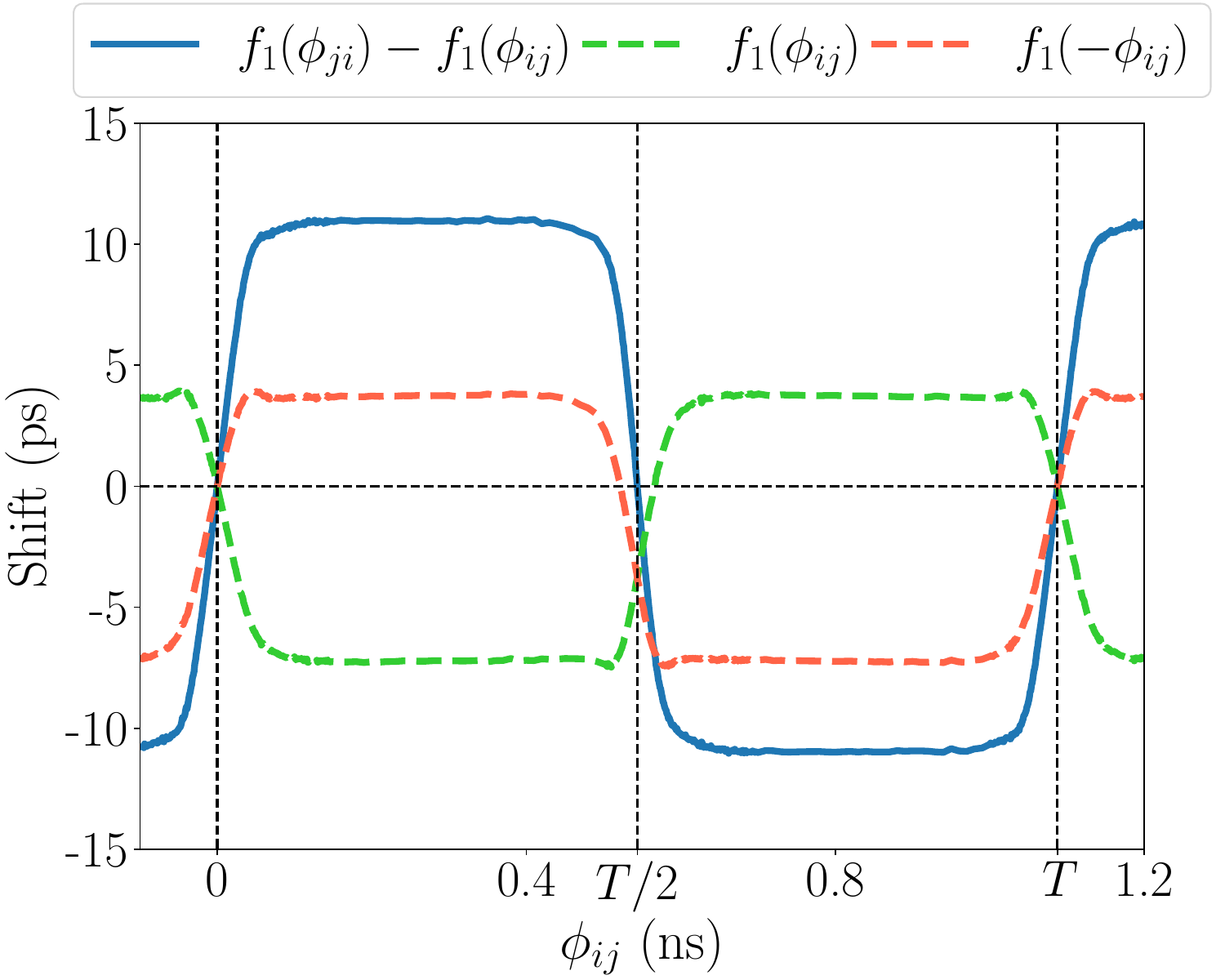}
    \label{fig:char_function}
    }
    \hspace{-2mm}
    \subfigure[]{\includegraphics[width=0.45\linewidth]{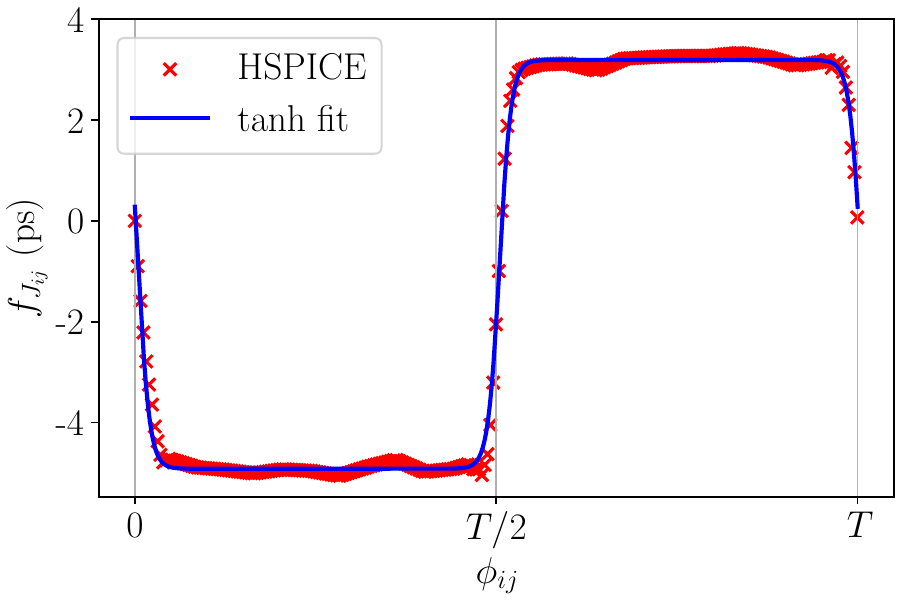}
    \label{fig:genAdler_vs_HSPICE}
    }
    \vspace{-3mm}
    \caption{(a)~The results for an example characterization setup, with $J_{ij}=1$, showing the delay shift $f_{J_{ij}}$ as a function of $\phi_{ij}$. (b)~HSPICE characterization of $f_{J_{ij}}$ and its tanh approximation as a function of phase difference $\phi_{ij}$.}
    \vspace{-2mm}
\end{figure}

\noindent
\underline{Traditional continuous-time formulations.}
The behavior of an LC oscillator, when injected with a sinusoidal signal, was described by Adler's equation~\cite{Adler}; a slightly different equation is used by Kuramoto~\cite{Kuramoto1984-hj}. To extend this beyond a single coupling and sinusoidal signals, the generalized Adler (GenAdler) equation for a network of $N$ coupled oscillators was shown~\cite{Bhansali2009} to have the form:
\begin{equation} \label{eq:gen-adler}
    \frac{d \phi_{i}(t)}{dt} = (\omega_i - \omega^*) + \omega_i \textstyle \sum_{j=1,j \neq i}^{N}c_{ij} (\phase{ij}{}(t)) 
\end{equation}
Here, $\phi_{ij}(t) = \phi_{i}(t) - \phi_{j}(t)$ is the difference between the phases of oscillators $i$ and $j$, $\omega_i$ is the frequency of the $i^{\rm th}$ oscillator, $\omega^*$ is the central frequency of the network, and for oscillators $i$ and $j$, $c_{ij}(.)$ is a $2\pi$-periodic function that represents the coupling-induced delay shift in each RO cycle. Prior methods~\cite{Bhansali2009} abstract $c_{ij}$ as a well-behaved function of $\phi_{ij}$, the phase difference of the coupled ring oscillators; Fig.~\ref{fig:char_function} shows an example HSPICE-characterized function showing the RO period shift against the phase difference.

\noindent
\underline{Discrete-time formulation.}
The continuous formulation models a discrete-event system composed of a sequence of coupling events.  We now examine the limitations of the continuous-time GenAdler model, as well as those of the phase delay shift model.

Fig.~\ref{fig:char_function} shows a model for the delay shift, $f_{J_{ij}}(\phi_{ij})$ of RO$_i$ and RO$_j$ at an example coupling strength $J_{ij} = 1$, where $\phi_{ij}$ is the phase difference between the two ROs. The delay shifts of RO$_i$ and RO$_j$ are shown by the red and green dotted lines, and the relative phase shift, i.e., the difference between these delay shifts, is shown by the solid blue line. It is seen that when the edges align to be in-phase or out-of-phase (i.e., at 0 or $T/2)$, the relative phase shift is zero.

We model the coupled system using a sequence of discrete events that add up to a shift in an RO clock period at the end of a cycle.  We denote the phase, period, and frequency as $\phase{i}{k}$, $\period{i}{k}$, and $\freq{i}{k}$, respectively, for RO$_i$ at the end of the $k^{\rm th}$ cycle. We use a datum oscillator frequency, $\omega^*$, which may be the frequency corresponding to the period $T$ of each uncoupled oscillator.

The phase shift of each oscillator from $\phase{i}{k}$ to $\phase{i}{k+1}$ during the $(k+1)^{\rm th}$ cycle is caused by two factors:\\
(1) frequency drift with respect to the reference oscillator:
\begin{align}
    \Delta \phase[1]{i}{k+1} = \phase[1]{i}{k+1} - \phase[1]{i}{k} 
                             = (\freq{i}{k} - \omega^*) \period{i}{k} 
    \label{eq:delta_phase_i_1} 
\end{align}
(2) phase/frequency drift due to coupling to other ROs:
\begin{align}
    \Delta \phase[2]{i}{k+1} = \phase[2]{i}{k+1} - \phase[2]{i}{k} 
                             = \textstyle \sum_{(i,j)\in E} f_{J_{ij}} (\phase{ij}{k}) 
    \label{eq:delta_phase_i_2}
\end{align}
where $E$ is the set of edges in the coupling graph (Section~\ref{sec:Background}-\ref{subsec:ro_ising}).  The net phase shift in the $(k+1)^{\rm th}$ cycle is
\begin{align}
\Delta \phase{i}{k+1} 
    =\Delta \phase[1]{i}{k+1} + \Delta \phase[2]{i}{k+1} 
    =(\freq{i}{k} - \omega^*) \period{i}{k} + \textstyle \sum_{(i,j)\in E} f_{J_{ij}} (\phase{ij}{k})
    \label{eq:delta_phase_i}
\end{align}
At synchronization, the clock frequency, $\omega_i^k$ is the same for all oscillators. Thus, in writing the relative phase difference between coupled oscillators RO$_i$ and RO$_j$ (note that $\phi_{ji} = -\phi_{ij}$), their corresponding first terms in~\eqref{eq:delta_phase_i} cancel, and we have:
\begin{align}
\Delta \phase{i}{k+1} - \Delta \phase{j}{k+1} = \textstyle \sum_{(i,j)\in E} f_{J_{ij}} (\phase{ij}{k}) - f_{J_{ij}}(-\phase{ij}{k}) = 0
\end{align}
The last equality arises because when the phases are locked, the difference between the delay shifts of locked oscillators is zero.

\noindent
\underline{Relation between the continuous- and discrete-time formulations.}
Unlike coupled sinusoidal oscillators, as modeled by the GenAdler formulation, where the signal value changes throughout a cycle, for an RO system, the coupling component of $d\phi_{i}(t)/dt$ changes during signal transitions but not when signals are stable at logic 0 or logic 1. 
Under infinitesimal phase changes per cycle, the derivative can be approximated if the net phase change over $m$ cycles is small. If the period of RO$_i$ in cycle $k$ is $\period{i}{k}$,  
\begin{align}
\frac{d\phi_{i}(t)}{dt} \approx \frac{1}{m} \frac{\sum_{k=1}^m \Delta\phase{i}{k}}{\Delta t},
\mbox{  where  } \Delta t = \textstyle \sum_{k=1}^m \period{i}{k}
\end{align}
From~\eqref{eq:delta_phase_i}, the phase change in time $\Delta t$ ($m$ cycles of RO$_i$) is 
\begin{align}
    \Delta \phi_{i} = \textstyle \sum_{k=1}^{m} \Delta \phase{i}{k} 
    =\textstyle \sum_{k=1}^m \left(\freq{i}{k} - \omega^* \right) T_i^k
        + \textstyle \sum_{(i,j)\in E} \textstyle \sum_{k=1}^{m} f_{J_{ij}} (\phase{ij}{k})         
    \label{eq:delta_phase_i_m_cycles}
\end{align}
The delay shifts over the $m$ cycles must be assumed to be small, i.e., $\freq{i}{k} \approx \omega_i \; \forall \; k = 1, \cdots, m$. Under this assumption, 
\begin{align}
\sum_{k=1}^{m} (\freq{i}{k} \period{i}{k} - \omega^*\period{i}{k}) 
    \approx (\omega_{i} - \omega^*) \sum_{k=1}^{m} \period{i}{k} 
    = (\omega_{i} - \omega^*) \Delta t
    \label{eq:summed_first_term}
\end{align}
Let $T_i$ be the average value of $\period{i}{k}$. Then  
\begin{align}
T_i =  \frac{1}{m} {\textstyle \sum_{k=1}^{m} \period{i}{k}}= \frac{\Delta t}{m}
\Rightarrow
m = \frac{\Delta t}{T_i} = \left( \frac{\omega_i}{2\pi} \right) \Delta t \label{eq:omega_to_m}
\end{align}
where $\omega_i \stackrel{\Delta}{=} 2\pi/T_i$.
Under small phase changes over $m$ cycles,  for $l = 1, 2, \cdots, k$, we write
$f_{J_{ij}} (\phase{ij}{l}) \approx f_{J_{ij}} (\phase{ij}{1}) + S_{ij}(\phase{ij}{l}-\phase{ij}{1})$ as a linear approximation,
where the sensitivity $S_{ij} = \left . \left ( \partial f_{J_{ij}}/\partial \phi_{ij} \right ) \right|_{\phase{ij}{1}}$. Therefore,
\begin{align}
\textstyle \sum_{k=1}^{m} f_{J_{ij}} (\phase{ij}{l}) 
    &= m \left[ f_{J_{ij}} (\phase{ij}{1}) + S_{ij} \left(\frac{\textstyle \sum_{k=1}^{m}\phase{ij}{k}}{m}  - \phase{ij}{1} \right) \right]
    \\
    &= \omega_i \left[ \frac{f_{J_{ij}} (\phi_{ij})}{2\pi}  \right] \Delta t \mbox{\hspace{4mm} (using \eqref{eq:omega_to_m})} \label{eq:summed_second_term}
\end{align}
where $\phi_{ij} = \sum_{k=1}^{m}\phase{ij}{k}/m$ is the mean value of $\phase{ij}{k}$ over $m$ cycles. 
From~\eqref{eq:delta_phase_i_m_cycles},~\eqref{eq:summed_first_term}, and~\eqref{eq:summed_second_term},
since $\Delta t$ is assumed to be small,
\begin{align}
    \frac{\partial \phi_i}{\partial t} &\approx \frac{\Delta \phi_i }{\Delta t} = (\omega_{i} - \omega^*) + \omega_i \sum_{(i,j)\in E}\frac{f_{J_{ij}} (\phi_{ij})}{2\pi} 
    \label{eq:final}
\end{align}
Setting $c_{ij} = f_{J_{ij}} (\phi_{ij})/(2\pi)$, this is the GenAdler equation,~\eqref{eq:gen-adler}.

\section{Limitations of the continuous-time approximation}
\label{sec:limitations_of_ct_approx}

\noindent
The continuous-time approach is effective in matching a coupled CMOS RO system when: 
\begin{enumerate}[label=(\alph*),noitemsep,topsep=-1pt,leftmargin=*]
    \item the phase difference between any pair of oscillators is independent of the location of coupling, and
    \item the phase difference between any pair of oscillators within a cycle is independent of their coupling to other oscillators. 
\end{enumerate}
From Section~\ref{sec:Adler_relation}, the function $c_{ij}(.)$ is crucial to the correctness of the model.  For CMOS ROs, coupling is expressed through complex MOS models (e.g., BSIM4, BSIM-CMG), the mapping from the system to the coefficient is nontrivial. 

\begin{figure}[htb]
    \centering
    \subfigure[]{\includegraphics[width=0.5\linewidth]{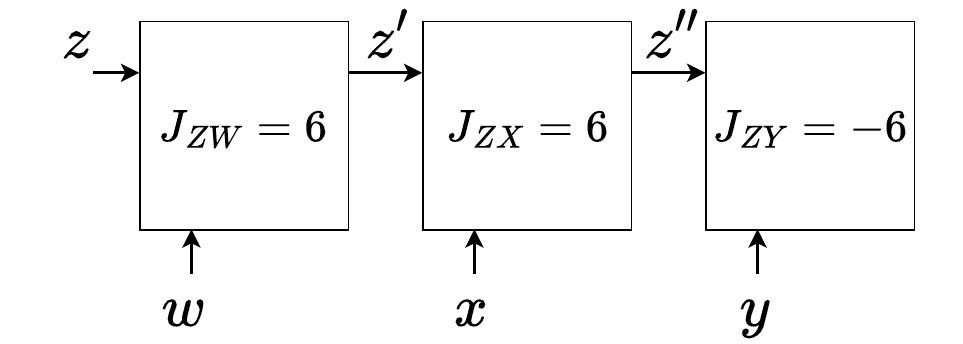}
    \label{fig:coupling_row}}
    
    \subfigure[]{\includegraphics[width=0.47\linewidth]{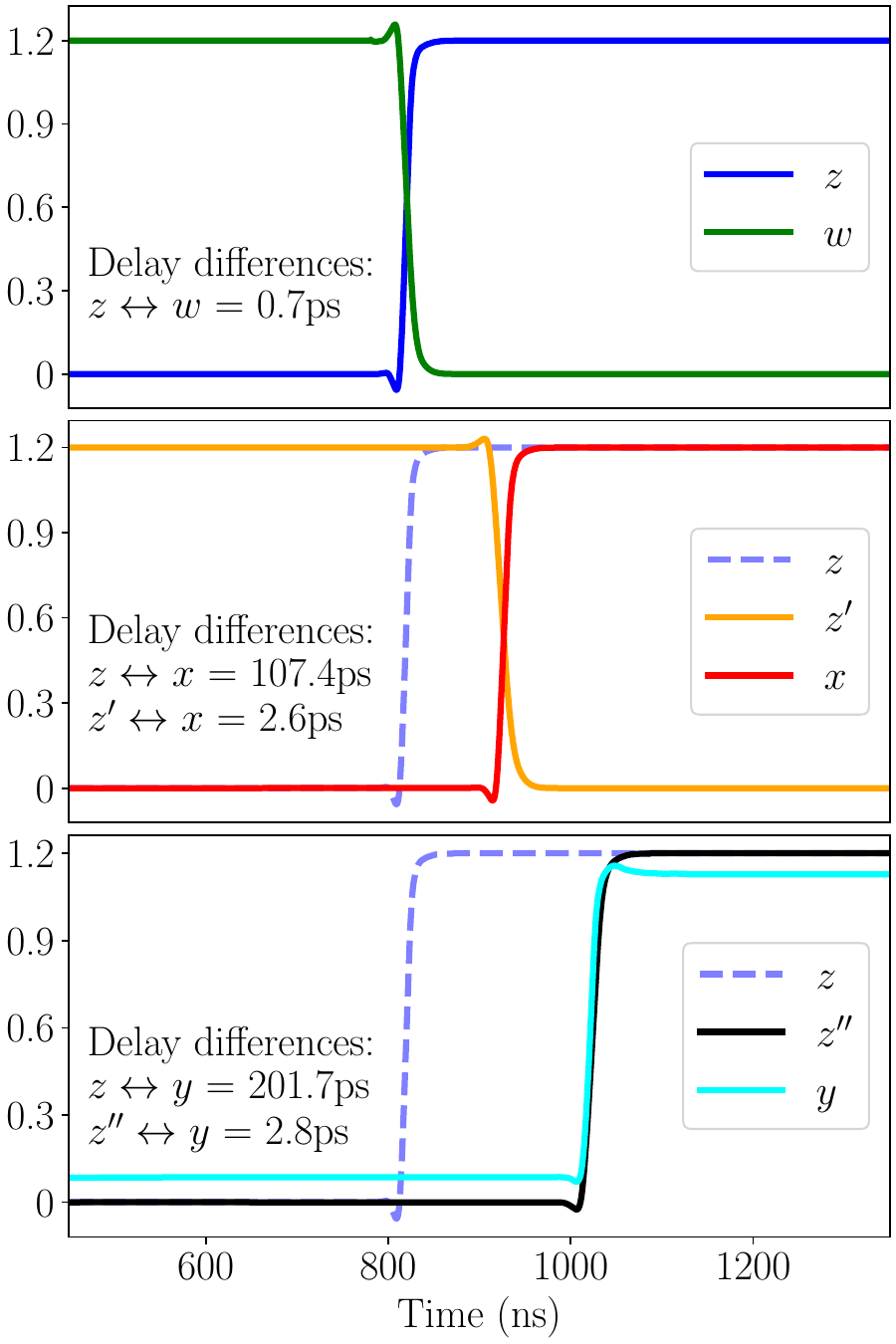}
    \label{fig:coupling_effect_base}}
    \subfigure[]{\includegraphics[width=0.47\linewidth]{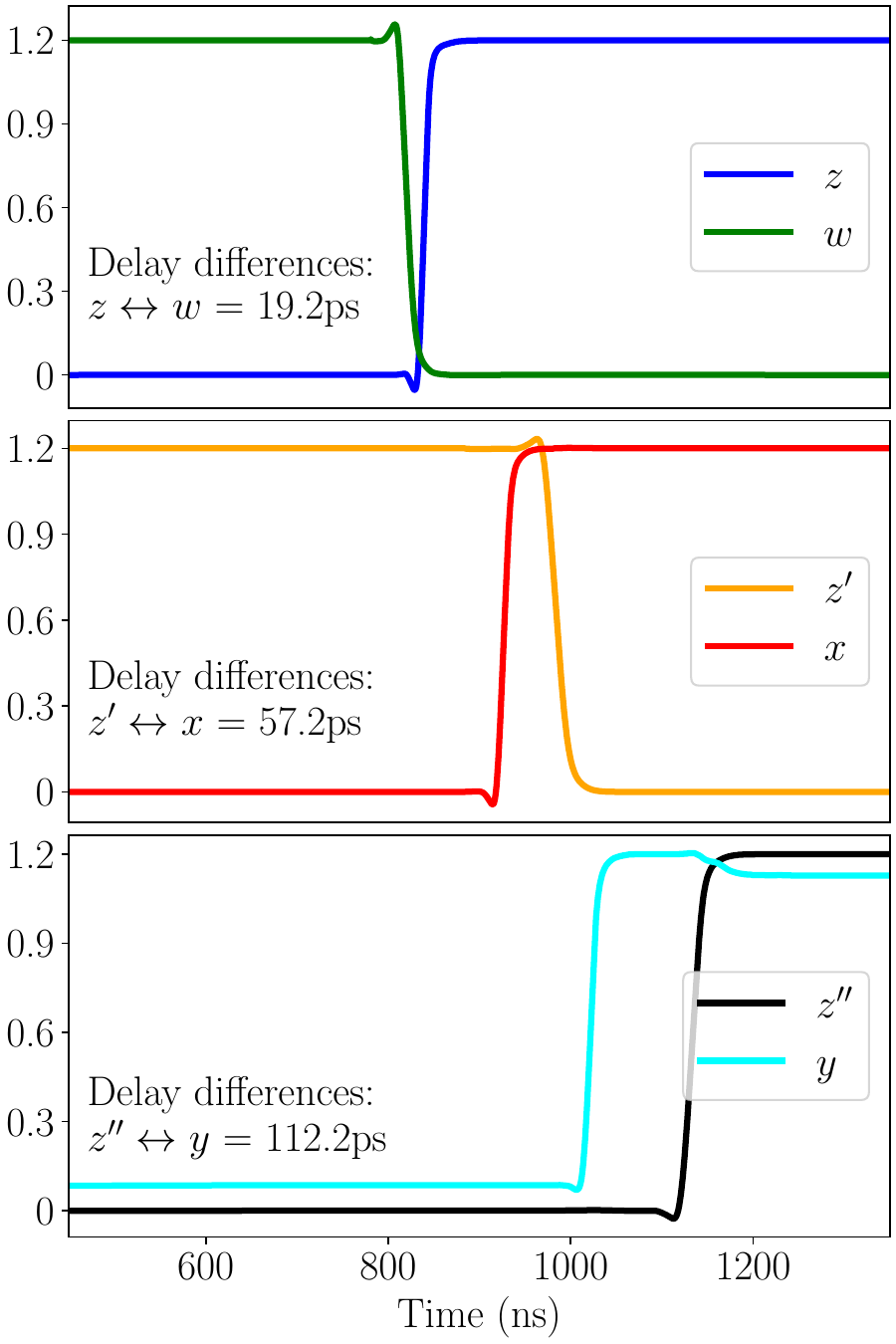}
    \label{fig:coupling_effect_shifted}}
    \vspace{-3mm}
    \caption{(a)~A section of the A2A array with three coupling cells. (b)~Phase difference at coupling cells depends on the delay difference at the inputs of the coupling cell and varies with location. (c)~The delay introduced by previous coupling cells affects the phase differences at later stages within the same cycle. 
    }
    \label{fig:coupling_example}
    \vspace{-4mm}
\end{figure}

To describe the impact of assumption~(a), we present an example that shows that the phase difference at a coupling cell in an A2A array depends on the location of the cell within the array and the magnitude of coupling in previous stages of the ROs. Our example in Fig.~\ref{fig:coupling_row} shows a section of an A2A array, where the labeled nets are inputs to a set of coupled cells in the array. The horizontal RO, RO$_Z$, runs through the lines $z$, $z'$, and $z''$. This oscillator couples with the vertical oscillators, RO$_W$, RO$_X$, and RO$_Y$, at the stages with inputs $w$, $x$, and $y$, respectively,  through tiles that implement the coupling coefficients, $J_{ZW} = 6$, $J_{ZX} = 6$, and $J_{ZY} = -6$. 
The locations $w$, $x$, $y$, and $z$ correspond to reference phases of respective ROs. Given a set of initial reference phases for each RO, we show the transitions at various locations in Fig.~\ref{fig:coupling_effect_base}. As mentioned in Section~\ref{sec:Background}-\ref{sec:practical_considerations}, the phase difference at the coupling cell sites may differ from the phase difference at the reference stages of their oscillators. For instance, the phase difference at $J_{ZX}$ is 2.6ps, corresponding to the difference in arrival times of transitions at $z'$ and $x$, which is different from the difference or 107.4ps in the arrival times of transitions at $z$ and $x$. As can be seen from Fig.~\ref{fig:transitions}, an inaccuracy of this magnitude (comparable to window $W$) will impact delay calculation for a coupling cell, also affecting the phase difference at the next cell.

Next, we consider the impact of assumption (b) alone, using corrected arrival times to eliminate the contribution of errors from assumption (a).  
As discussed in Section~\ref{sec:Background}-\ref{sec:practical_considerations}, RO delays can vary with the arrival time difference, and these delays are also subtly impacted by changes in the signal transition time at the RO input.
We determine the coupling between oscillators in Fig.~\ref{fig:coupling_effect_shifted} for a case where the arrival times of transitions at $w$, $x$, and $y$ are identical, but the transition at $z$ is slightly delayed, compared to the corresponding values in Fig.~\ref{fig:coupling_effect_base}.  This small change has the effect of changing the coupling delay and transition time at $z'$, with a ripple effect on the timing of transitions at $z'$ and $z''$, caused by the coupling between RO$_Z$ and other oscillators: it can be seen that the small shift of $<$20ps at $z$ shifts the transition at $z'$ by $>$50ps, and at $z''$ by $>$100ps.
These magnified shifts arise because the arrival time at $z'$ depends on the magnitude of coupling, $J_{ZW}$, and that at $z''$ depends on both $J_{ZW}$ and $J_{ZY}$. Thus, it is not just the number of stages from the reference to the coupling stage that affects the delay shift; the magnitude of coupling in previous stages, and the precise timing relationship between waveforms in those stages, also affect phase differences at a later stage.

The GenAdler formulation in~\eqref{eq:gen-adler} makes assumptions (a) and (b), and simply uses the phase difference $\phi_{ij}$ between the reference stages of oscillators $i$ and $j$.  In Section~\ref{sec:Simulation}, we present a fine-grained event-driven approach to overcome these limitations. 

We know of only one prior event-driven approach~\cite{sreedhara23}, but it uses a fundamentally different definition of events from ours, and speeds up the generalized Kuramoto simulation. This method inherits the assumptions as GenAdler, and hence, limitations (a) and (b).  Its speedup mechanism determines whether, at any time, two or more ROs achieve a phase difference that is an integral multiple of $\pi$ radians: if so, it assumes that these oscillators will remain permanently phase-locked from that time onwards. 
Through this assumption, the number of variables is reduced as the simulation proceeds, reducing its computational cost.

\begin{figure}[htb]
    \centering
    \includegraphics[width=0.95\linewidth]{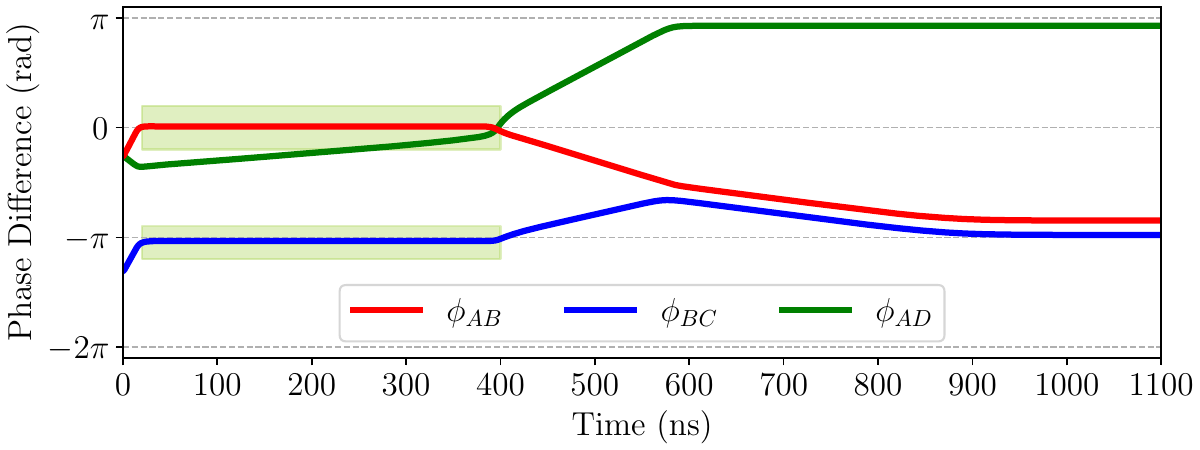}
    \vspace{-4mm}
    \caption{The phase differences across various edges are shown above. The phase differences $\phi_{AB}$ (red) and $\phi_{BC}$ (blue) stay close to $0$ and $\pi$ between 20ns to 400ns giving the impression of being phase-locked. The phase differences deviate from these seemingly phase-locked positions beyond 400ns. 
    }
    \label{fig:diverge}
    \vspace{-8mm}
\end{figure}

We show a counterexample, based on HSPICE simulations, that illustrates that such preliminary phase-locking assumptions~\cite{sreedhara23} can be incorrect.  Consider a system comprising four ROs, denoted as RO$_A$, RO$_B$, RO$_C$, and RO$_D$. Fig.~\ref{fig:diverge} shows the phase differences in radians between coupled RO pairs as the system evolves in time. The phase difference $\phi_{AB}$ (red) remains close to 0 from 20ns to 400ns and $\phi_{BC}$ (blue) remains close to $-\pi$ from 20ns to 400ns, as shown in the highlighted green box. In the interval [20ns, 400ns], it appears as if RO$_A$ and RO$_B$ are locked in-phase while RO$_B$ and RO$_C$ are locked out-of-phase, but oscillator RO$_D$ is not yet settled as $\phi_{AD}$ (green) continues to change. As shown in the example in Fig.~\ref{fig:coupling_example}, couplings in earlier stages affect delays within the same cycle, and it is a result of this effect that the changes in $\phi_{AD}$ become more dramatic around $t=$ 400ns, it causes phase differences at other coupling cells to change. The net effect of these changes on RO$_A$, RO$_B$, and RO$_C$ is that they leave their seemingly phase-locked relationships as the system evolves, and settle to a different equilibrium at $t=$ 900ns, where RO$_A$ is out-of-phase with RO$_B$, and RO$_D$, and in-phase with RO$_C$. If the phases of RO$_A$ and RO$_B$ (or RO$_B$ and RO$_C$) were merged into a single phase based on their behavior between 20ns and 400ns, this equilibrium stage would not be captured by the simulation.

%% file: sec/4_Simulation.tex
\section{Simulating the A2A array}
\label{sec:Simulation}

\subsection{Capturing the timing information}
\label{subsec:timing_capture}

\noindent 
We define an \textit{event} as a rise or fall transition in a digital signal at the input of a logic gate in an RO, which can cause the opposite transition at its output. We characterize each transition by its arrival time, transition time (rise/fall time), and whether the signal is rising or falling. Our approach is motivated by timing analysis in CMOS digital design, where a timing arc is used to propagate an event at the input of a gate to an event at its output.  
Cell timing information, i.e., the input-to-output delay and the output transition time, is captured in lookup tables as functions of the output load and the transition time of the input signal. 

The invariant during timing analysis is the computation of the arrival time and transition time at a node. 
Given an event at the input of a cell, characterized by these two values, the timing information of the cell can be used to generate an output event(s) and their arrival time(s) and transition time(s).  These events are expressed at the input of another cell, and the process continues.

As mentioned in Section~\ref{sec:A2A}, the A2A array has three types of cells: enable, coupling, and shorting. The simulator works with a timing view of these cells, and in the remainder of this section, we discuss this timing abstraction, using the notation in Fig.~\ref{fig:array_schm}.

The {\em enable cell} has two inputs, one enable signal, \textit{enable}, and another from the RO itself (\textit{inp}). The cell is modeled by a timing arc from \textit{inp} to \textit{outp}.  Since the load is the same for all enable cells, a one-dimensional table, characterized using HSPICE, is used to represent the cell rise delay as a function of the input transition time; a similar table characterizes the fall delay.

The {\em coupling cell} has four inputs ($h_{in}^{f}$, $v_{in}^{f}$, $h_{in}^{r}$, and $v_{in}^{r}$) and four outputs ($h_{out}^{f}$, $v_{out}^{f}$, $h_{out}^{r}$, and $v_{out}^{r}$), where the symbols $h$ and $v$ correspond to the horizontal and vertical ROs, and the superscripts $f$ and $r$ represent the forward and reverse path, respectively, through the cell. The horizontal timing arc ($h_{in}^{f}$ to $h_{out}^{f}$) of the horizontal RO interacts with the vertical timing arc ($v_{in}^{f}$ to $v_{out}^{f}$) of the vertical RO within a window when the cell implements a non-zero coupling coefficient; otherwise the horizontal and vertical paths through the cell do not interact. 
To represent timing on the forward path, we use HSPICE-characterized three-dimensional tables for the delay and output transition times, indexed by the transition times of the two inputs, $h_{in}^{f}$ and $v_{in}^{f}$, and the difference between the arrival times of the two input events, which ranges from $-W$ to $+W$. The precise value of the interaction window width, $W$, defined in Section~\ref{sec:Background}-\ref{sec:practical_considerations}, is determined from HSPICE simulations. Each input may rise or fall, and the four resulting combinations of the transition types imply that we require four tables per coupling value. As a coupling cell implements $2C_{max} + 1$ levels, a total of $4(2C_{max} + 1)$ three-dimensional tables are required.
The timing arcs for the return paths ($h_{in}^{r}$ to $h_{out}^{r}$ and $v_{in}^{r}$ to $v_{out}^{r}$) do not interact as they are not coupled in the A2A architecture of Section~\ref{sec:A2A}. Therefore, the events on these arcs can be processed independently of each other, and one-dimensional tables will suffice as in the case of the enable cell.

The {\em shorting cell} has four inputs and four outputs that are labeled in the same way as the coupling cell, and the difference is that the coupling between the horizontal and vertical oscillators here is a short circuit. Since both the horizontal and vertical oscillators that meet at a shorting cell $(i, i)$ are enabled by the same enable signal, $en_i$, any phase difference between them is a result of differences in coupling delays between the horizontal and vertical oscillators. Since Ising hardware uses weak coupling, these differential delays constitute a small fraction of the period. As a result, the arrival of a rising transition on the vertical RO will not be so severely delayed that it interacts with the falling transition of the horizontal RO at a shorting cell. Therefore, the lookup tables that capture rise-fall and fall-rise interactions are unnecessary, and two lookup tables suffice for shorting cells.

\subsection{Overview of the event-driven simulator}
\label{subsec:simulator}
\vspace{-2mm}
\noindent
The simulator requires the following inputs:
(1)~a \textit{timing file}, with the characterized lookup tables and the interaction window \mbox{(Section~\ref{sec:Simulation}-\ref{subsec:timing_capture})};
(2)~the \textit{circuit netlist}, a file that hierarchically captures the connections between devices and circuits;
(3)~a problem-specific \textit{coupling matrix}, which maps the coupling coefficients of the Hamiltonian to the coupling cells, and is used to select the appropriate lookup table during simulation; 
(4)~a \textit{maximum simulation time}, which specifies the total simulation time; and
(5)~a \textit{tolerance} value used to check for RO synchronization (Section~\ref{sec:Background}-\ref{subsec:ro_ising}).
We use the following data structures:
\begin{itemize}[noitemsep,topsep=-1pt,leftmargin=*]
    \item \textbf{Event:}  
    an object that records an event, recording the net name, arrival time, transition time, and transition type (rising or falling).
    \item \textbf{Q:} a queue that sorts events by their arrival time, with the earliest occurring event at the head.
    \item \textbf{Net2Event:} a map with a net name as the key, pointing to an event at that net.
    \item \textbf{PendingTrigger:} a map with a net name as the key, pointing to a pending event with insufficient information for processing.
\end{itemize}

\begin{figure}[H]
    \centering
    \includegraphics[width=0.5\linewidth]{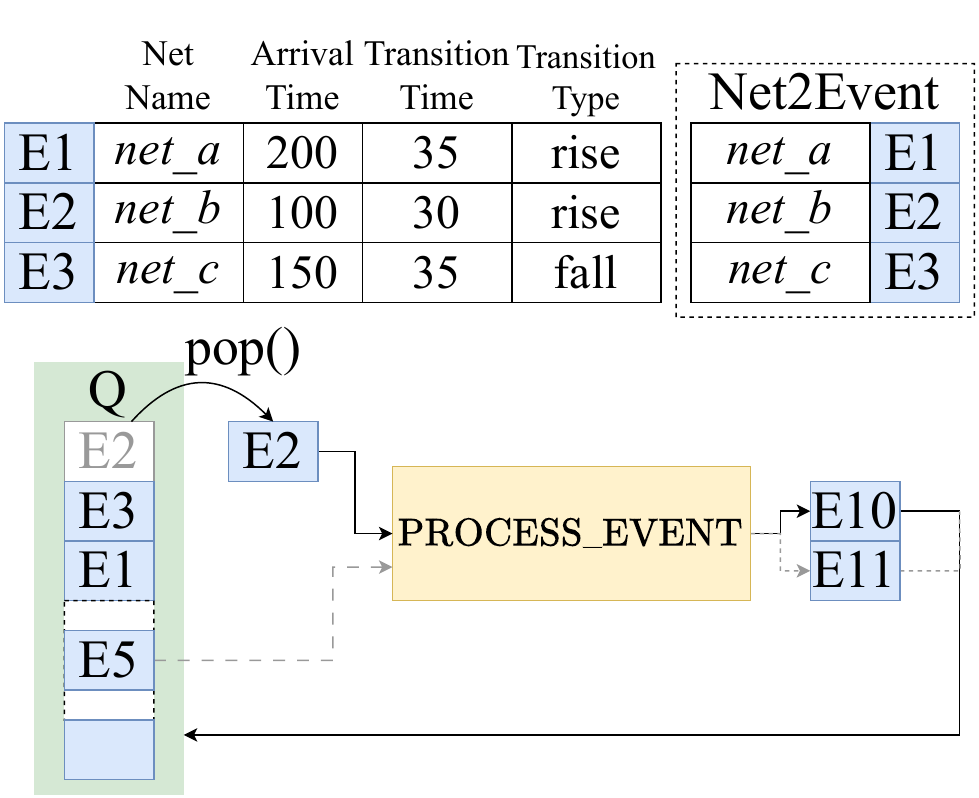}
    \vspace{-2mm}
    \caption{A simulator step with the sorted queue Q and the Net2Event map: \textproc{process\_event} consumes one or more events and generates future events.}
    \label{fig:overview}
    \vspace{-2mm}
\end{figure}

\noindent
The simulator outputs are the map Net2Event and \textit{spin\_vals}, the set of spins that optimize the Hamiltonian for the \textit{coupling matrix}.

An overview of the simulator is shown in Fig.~\ref{fig:overview}, listing the event objects, the map Net2Event, and the scheduled events in queue Q. The simulator algorithm, described by pseudocode in Algorithm~\ref{alg:sim}, consists of the following steps:

\input{pseudocode/topAlgo}

\noindent
\textbf{Step 1: Initialize} Initial events at the enable cells are scheduled to start the ROs. The queue, Q, and the map, Net2Event, are populated to reflect these events, and PendingTrigger is initialized to an empty map.

\noindent
\textbf{Step 2: Pop and process an event} The earliest occurring event E is popped from Q. The event is passed to the \textproc{process\_event} function which generates new events that result from E. Consider an event that occurs at the $v^f_{in}$ pin of a coupling cell and the map Net2Event contains another event that occurs at $h^f_{in}$ of the same cell. Then, \textproc{process\_event} will operate on these two events to generate events on output pins $v^f_{out}$ and $h^f_{out}$, of the coupling cell. We describe the \textproc{process\_event} function in Section~\ref{sec:Simulation}-\ref{subsec:process_event}.

\noindent
\textbf{Step 3: Check timeout and synchronization criteria} The timeout criterion is met if the earliest event scheduled in Q has exceeded the \textit{maximum simulation time}.
The synchronization criterion, as defined in Section~\ref{sec:Background}-\ref{subsec:ro_ising}, is met when the periods of all coupled ROs are within the specified \textit{tolerance}. We terminate the simulation when either of the above criteria is met.

\noindent
\textbf{Step 4: Assign spin values} At the end of the simulation, the RO phases are translated to spin values, assigning a spin of $+1$ to the reference RO. The phase difference between the RO in the A2A array and every other RO in the array is determined: if this phase difference is closer to $0$ than it is to $\pi$, a spin value of $+1$ is assigned to the RO, otherwise, we assign a spin value of $-1$.

\begin{figure*}[!ht]
{\centering
\subfigure[]{\includegraphics[width=0.32\linewidth]{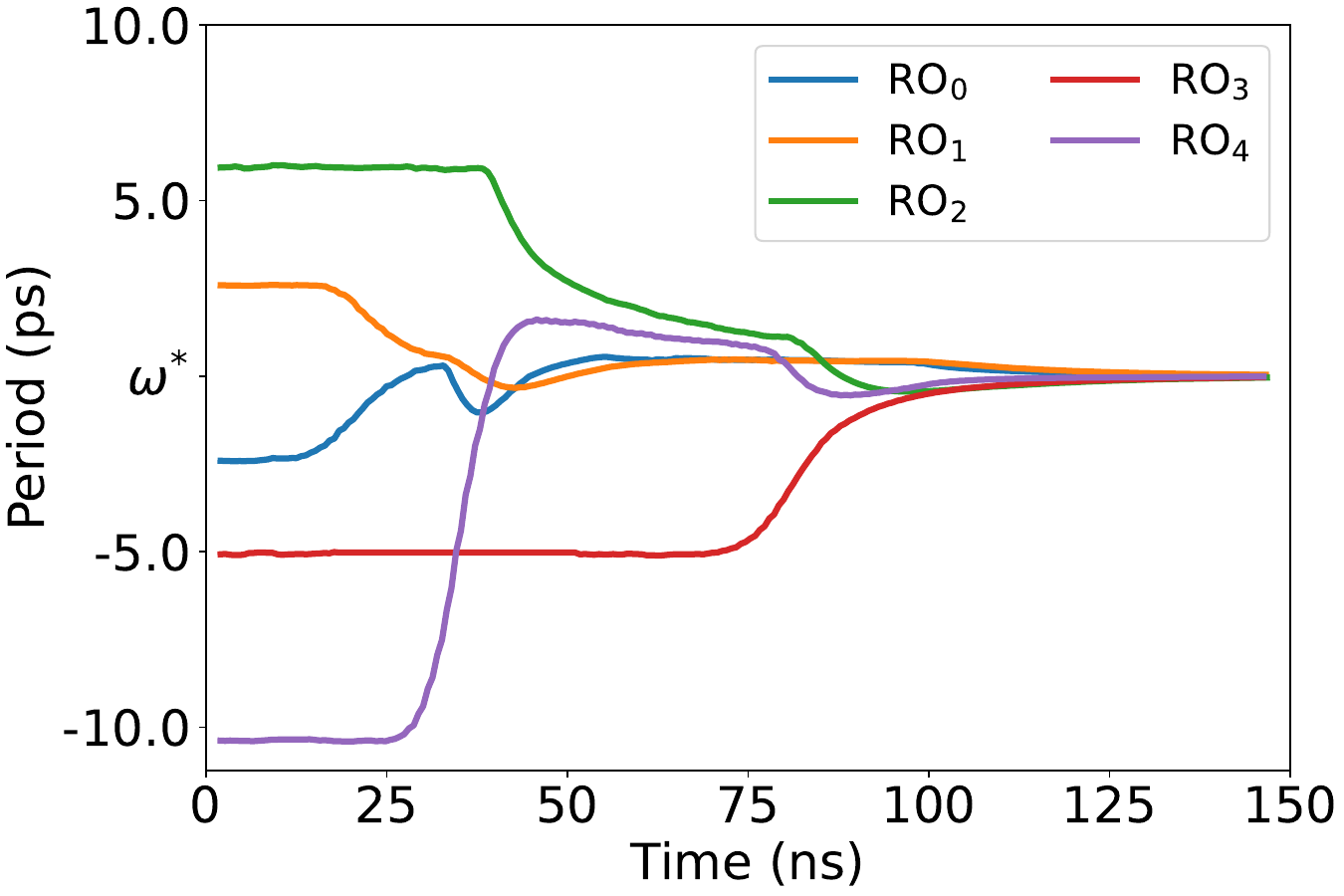}
\label{fig:droid_wave}}
\subfigure[]{\includegraphics[width=0.32\linewidth]{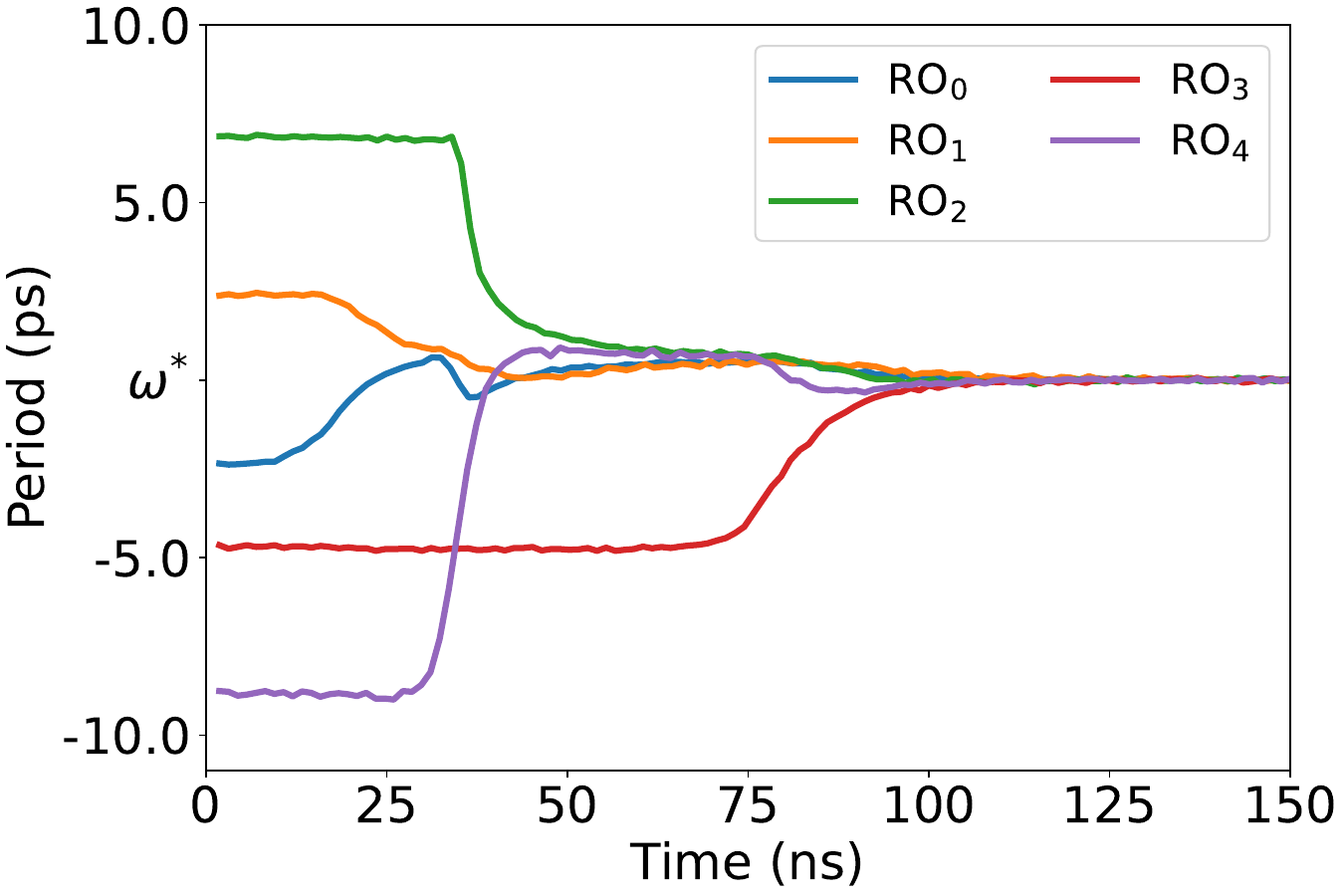}
\label{fig:hspice_wave}}
\subfigure[]{\includegraphics[width=0.32\linewidth]{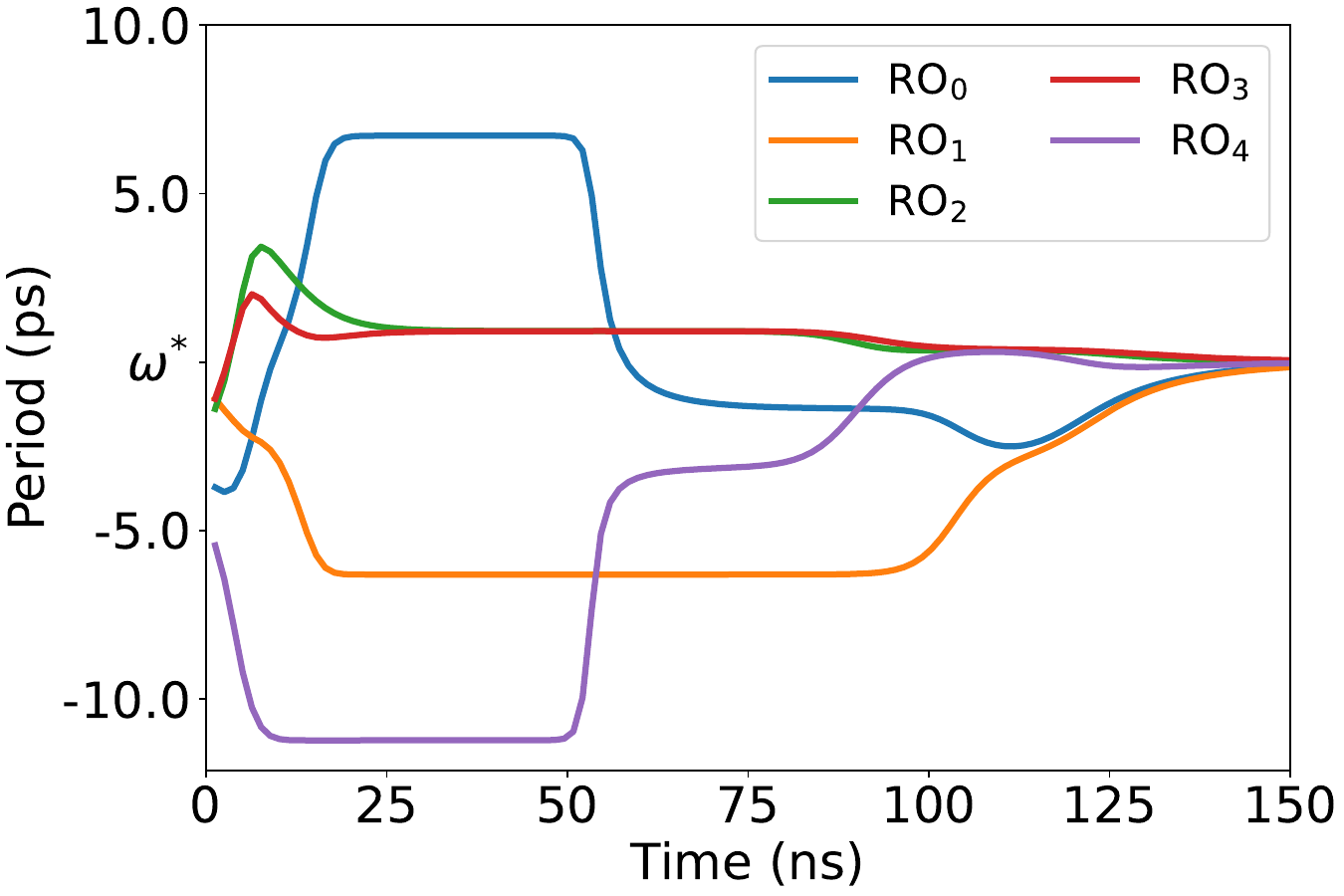}
\hfill
\label{fig:genAdler_wave}}
}
\caption{Period waveforms for a {5$ \times $5} A2A array from (a)~DROID, (b)~HSPICE, and (c)~GenAdler for the same initial conditions of the ROs.}
\label{fig:waveforms_comparison}
\vspace{-6mm}
\end{figure*} 

\subsection{\textproc{process\_event}: Processing an event from the queue}
\label{subsec:process_event}

\noindent
We describe the intuition behind \textproc{process\_event} using an example to convey the complexities of looking forwards and backwards in time within the interaction window $W$; the pseudocode for \textproc{process\_event} is provided in Appendix~\ref{app:appendix}-\ref{app:process_event}.

\begin{figure}[H]
    \centering
    \includegraphics[width=0.8\linewidth]{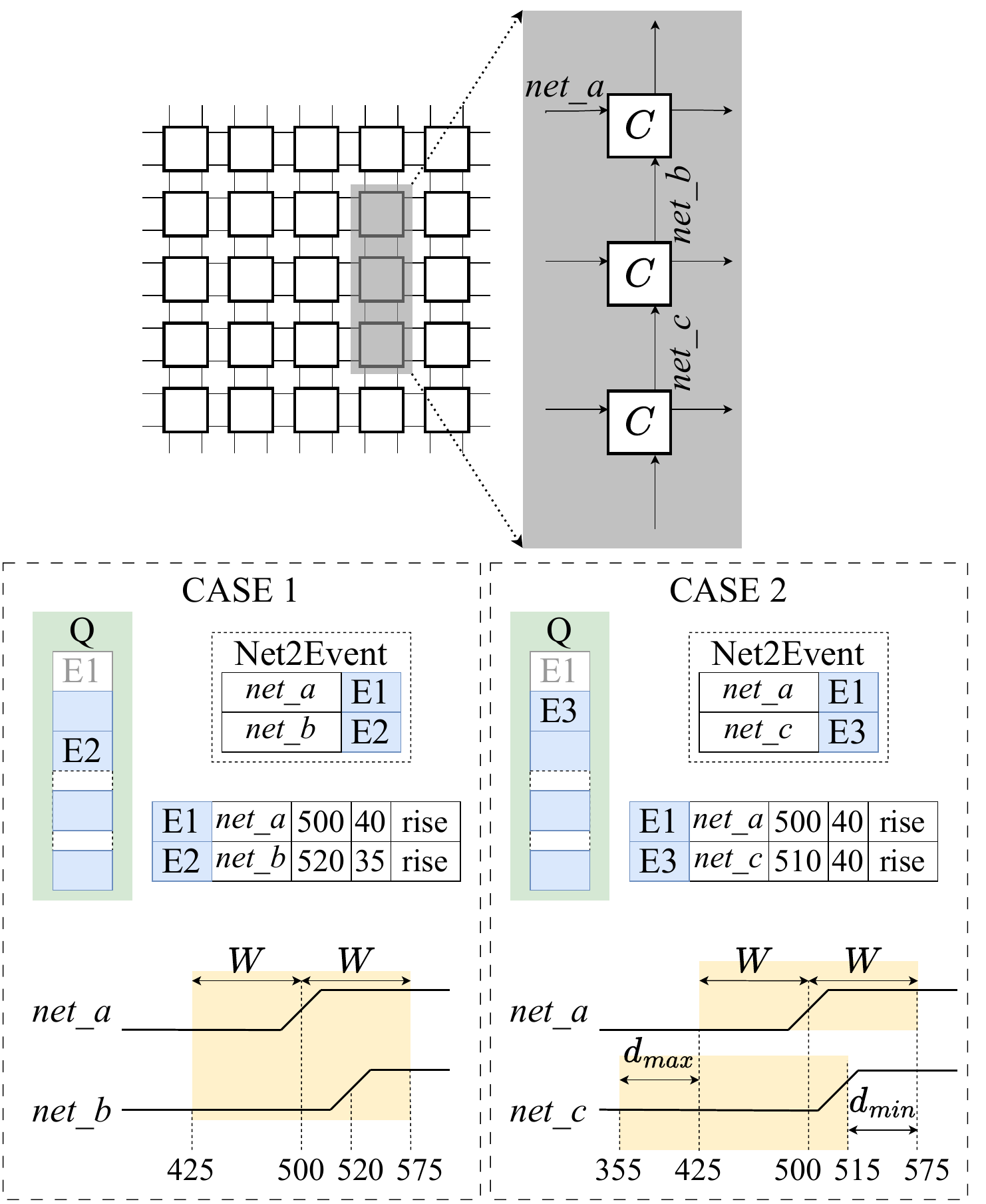}
    \caption{The handling of an event on \textit{net\_a} is influenced by the knowledge of events on nets in some neighborhood around it, as shown on the top. \mbox{\uppercase{Case 1}} shows the scenario when an event on the other input (\textit{net\_b}) of the same instance is known. \uppercase{Case 2} shows the scenario where \textproc{look\_back} is invoked to find an event on \textit{net\_c} that can cause an event on \textit{net\_b} that might lie within the interaction window of the event at \textit{net\_a}.
    }
    \label{fig:process_event}
    \vspace{-4mm}
\end{figure}

\noindent
\textbf{Example 1:} 
Fig.~\ref{fig:process_event} shows a $5 \times 5$ A2A array, and focuses on three coupling cells within the array, as shown in the inset. The figure depicts two separate scenarios, \uppercase{Case 1} and \uppercase{Case 2}, that will be used as examples in this subsection. We assume that the timing file specifies $W=$ 75ps for both examples. 
Consider the situation shown in \uppercase{Case 1} of Fig.~\ref{fig:process_event} where \textproc{process\_event} is called on a rising transition at \textit{net\_a}, which arrives at $t=$ 500ps and has a transition time of 40ps.
The Net2Event map shows a rising transition on \textit{net\_b} at 520ps, with a transition time of 35ps. 
As the arrival time difference of the events is 20ps which is less than the window, these events interact.

The output events are calculated using the three-dimensional lookup table mentioned in Section~\ref{sec:Simulation}-\ref{subsec:timing_capture}.  Note that if the event at \textit{net\_b} were to arrive at 580ps instead of 520ps, it would not interact with the event at \textit{net\_a}. In such a scenario, the event at \textit{net\_a} would be processed as a non-interacting event.  The event(s) generated from processing \textit{net\_a} are inserted into Q, and any key-value pairs in Net2Event associated with \textit{net\_a} and any interacting event are removed.
\hfill $\Box$

The above example considers events already in the Net2Event map, but the process could be complicated by as-yet-unprocessed events that could interact with a transition under consideration. For example, if \textit{net\_b} is not a key in Net2Event, \textproc{look\_back} is used to examine the predecessors of \textit{net\_b} to determine whether any upcoming event might interact with the event on \textit{net\_a}. We illustrate this with an example of a call to \textproc{look\_back}; the pseudocode for \textproc{look\_back} is provided in Appendix~\ref{app:appendix}-\ref{app:look_back}.

\noindent
\textbf{Example 2:} Consider \uppercase{Case 2} in Fig.~\ref{fig:process_event} with events at \textit{net\_a} and \textit{net\_c} in Q. To process the event at \textit{net\_a} which arrives at 500ps, an interacting event on \textit{net\_b} should arrive in the window (425ps, 575ps); there is no event in Net2Event with the key \textit{net\_b}. Thus, \textproc{process\_event} invokes \textproc{look\_back} with the arguments (\textit{net\_b}, (425ps, 575ps), 425ps, Net2Event). 
The predecessor of \textit{net\_b} is \textit{net\_c}.  
Assume for this example, that the minimum and maximum delays of the coupling cell obtained from the timing file are 60ps and 70ps, respectively.  
An event that occurs on \textit{net\_c} can occur as early as 355ps to incur the maximum delay of 70ps and still generate an event on \textit{net\_b} in the required window. Similarly, an event on \textit{net\_c} can occur as late as 515ps and incur the minimum delay of 60ps to generate an interacting event on \textit{net\_b}. Thus, the window of arrival for an event on \textit{net\_c} is (355ps, 515ps).

Since Net2Event contains an event on \textit{net\_c} within this window, \textproc{look\_back} returns true. In \textproc{process\_event}, we stall the processing of \textit{net\_a} until \textit{net\_b} is scheduled, by adding the event to the map PendingTrigger with a key \textit{net\_b}. When the event at \textit{net\_c} is processed and it generates another at \textit{net\_b}, the pending event on \textit{net\_a} will be added back to the queue.
\hfill $\Box$

%% file: pseudocode/topAlgo.tex
\begin{algorithm}[tb]
    {
    \small
    \caption{Simulation of an A2A array of ROs}
    \label{alg:sim}
    \begin{algorithmic}[1]
    \State \textbf{Input}: Timing file, circuit netlist, coupling matrix, maximum simulation time, and tolerance.  
    \State \textbf{Output}: A spin assignment for ROs.
    \State \textit{// Step 1: Initialize.}
    \State initial\_events $\xleftarrow{}$ Initialize with events on enable pins 
    \State Net2Event, PendingTrigger $\xleftarrow{}$ map()
    \For {event $\in$ initial\_events} 
        \State Q.add(event)
        \State Net2Event[event.netname]    = event
    \EndFor
    \State timeout $\xleftarrow{}$ False
    \State synchronized $\xleftarrow{}$ False
    \While {!timeout \textbf{and} !synchronized} 
        \State \textit{// Step 2: Pop an event for processing.}
        \State E $\xleftarrow{}$ Q.pop()
        \State \Call{process\_event}{E, Q, Net2Event, PendingTrigger}
        \State \textit{// Step 3: Check timeout and synchronization criteria.}
        \If {Q[0].arrival\_time $>$ \textit{maximum simulation time}}
            \State timeout $\xleftarrow{}$ True
        \EndIf
        \If {synchronization criteria met}
            \State \textit{// Synchronization condition defined in Section~II-C.}
            \State synchronized $\xleftarrow{}$ True
        \EndIf
    \EndWhile
    \State \textit{// Step 4: Assign spin values} 
    \State spin\_vals $\xleftarrow{}$ Assign spins based on the phase difference of each RO with the reference RO
    \State \Return spin\_vals, Net2Event
    \end{algorithmic}
    }
\end{algorithm}

%% file: sec/5_Results.tex
\section{Results and discussion}
\label{sec:Results}

\noindent
We have implemented DROID in Python 3.8.8 on a 64-bit RHEL 7.9 server with a 2.2GHz Intel\textsuperscript{\tiny\textregistered} Xeon\textsuperscript{\tiny\textregistered} Silver 4114 CPU. We used HSPICE for circuit simulation using models from a commercial low-power 65nm technology. The characterization to build timing models for the cells, as described in Section~\ref{sec:Simulation}-\ref{subsec:timing_capture}, took roughly 14 hours. This is a one-time step at a technology node, similar to standard-cell library characterization, and its results can be reused across all designs and all simulations in the technology node. Therefore, like standard-cell library characterization, this amortization makes the runtime acceptable. 

\noindent
\underline{Comparison of DROID with GenAdler and HSPICE}
Fig.~\ref{fig:waveforms_comparison} shows period waveforms of five ROs in a {5$\times$5} A2A array as predicted by DROID, GenAdler, and HSPICE, for the same initial conditions, showing the evolution of RO states.  The GenAdler simulation is based on available code~\cite{wang2019matlab} and uses the tanh approximation for $f_{J_{ij}}$.
In all plots, the period of each of the five ROs, labeled RO$_0$ through RO$_4$, evolves over the simulation until it reaches its final value.  As pointed out in Section~\ref{sec:Background}-\ref{subsec:ro_ising}, the system tries to synchronize to a common period after which the phase differences between ROs become constant.  It can be seen that the period waveform from DROID in Fig.~\ref{fig:droid_wave} matches that from HSPICE (Fig.~\ref{fig:hspice_wave}), but the waveform from  GenAdler (Fig.~\ref{fig:genAdler_wave}) is noticeably different from HSPICE.
Inaccuracies in the GenAdler waveforms are attributed to the limitations discussed in Section~\ref{sec:limitations_of_ct_approx}. 

\noindent
\underline{Runtime trend as a function of array size}
We apply DROID to the silicon-proven A2A-coupled RO array~\cite{Lo2023} (Section~\ref{sec:A2A}). Table~\ref{tbl:runtime} compares the runtimes of DROID and HSPICE for various array sizes, simulated up to 100ns, for a single initial condition, over three array sizes, 5$\times$5, 20$\times$20, and 50$\times$50, for random Ising problems. The HSPICE runtime increases superlinearly with the array dimension, and is prohibitively large even for a single initial condition; in contrast, DROID is computationally efficient. 

\input{tables/simulation_runtimes}

The analysis above considers a single graph topology and a single initial condition on the graph. We examine the scalability of DROID by determining the runtime for simulating the solution of Ising problems on A2A arrays of sizes up to 50$\times$50, across various random graphs, and for multiple initial conditions.  Recall from Section~\ref{sec:A2A}, that an A2A array of size N$\times$N is equivalent to a hexagonal/King's graph array with $\sim$$N^2$ coupled ROs~\cite{Tabi21, Lucas2019}; therefore, a 50$\times$50 A2A array has equivalent computation power to a $\sim$2500-spin planar array, and avoids the problems of weakened spins caused by minor embedding in planar RO arrays.  The random graphs are characterized by the value of graph density, defined as $2|E|/(|V||V|-1)$, where $|V|$ is the number of vertices (i.e., ROs in the array) and $|E|$ is edge count: the density varies from 0 for a graph with no edges to 1 for a complete graph. 

\begin{figure}[htb]
    \centering
    \includegraphics[width=0.6\linewidth]{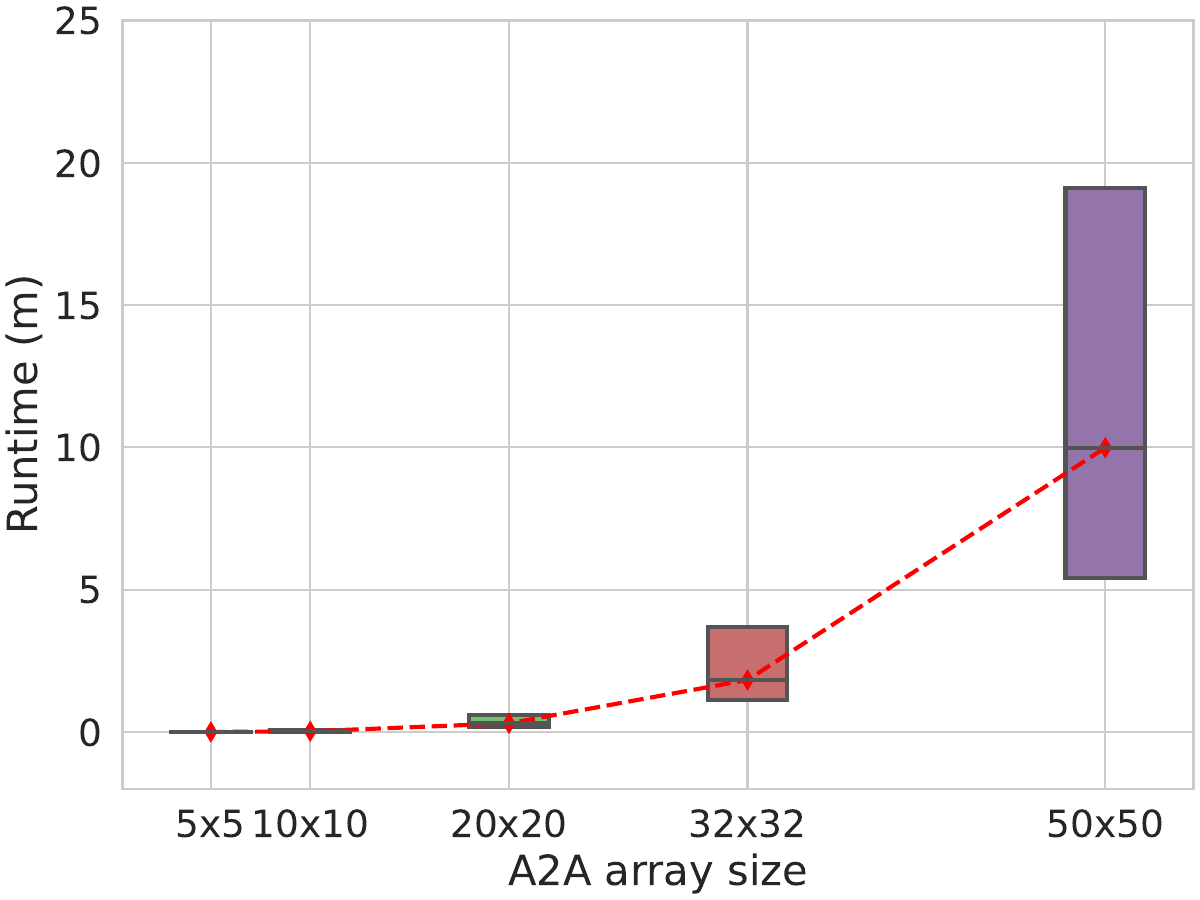}
    \vspace{-4mm}
    \caption{Distribution of DROID runtimes for various A2A array sizes.
    }
    \label{fig:runtime}
    \vspace{-6mm}
\end{figure}

DROID was exercised on arrays of various sizes: 5$\times$5, 10$\times$10, 20$\times$20, 32$\times$32, and 50$\times$50.  
For each size, we generate 80 random problems, 20 problems per graph density $\in \{0.4, 0.6, 0.8, 1.0\}$, with coupling values from a uniform probability distribution over $[-7, +7]-\{0\}$.
Each problem was simulated with different initial conditions up to synchronization with a maximum simulation time of 250$\mu$s. The distribution of DROID runtimes that attained synchronization for each array size is shown in Fig.~\ref{fig:runtime}. Each box shows the lower quartile, median, and upper quartile of the runtimes over more than 1000 simulations for each array size. 
As can be seen from the figure, the median runtime of synchronized runs for a 50$\times$50 array is 10 minutes and the upper quartile is 19.1 minutes. Such an experiment is not possible on HSPICE, since even a single run takes over 16 hours to simulate a 50$\times$50 array for 100ns, and 1000 simulations would take more than 20 months.

\begin{figure}[H]
    \centering
    \includegraphics[width=\linewidth]{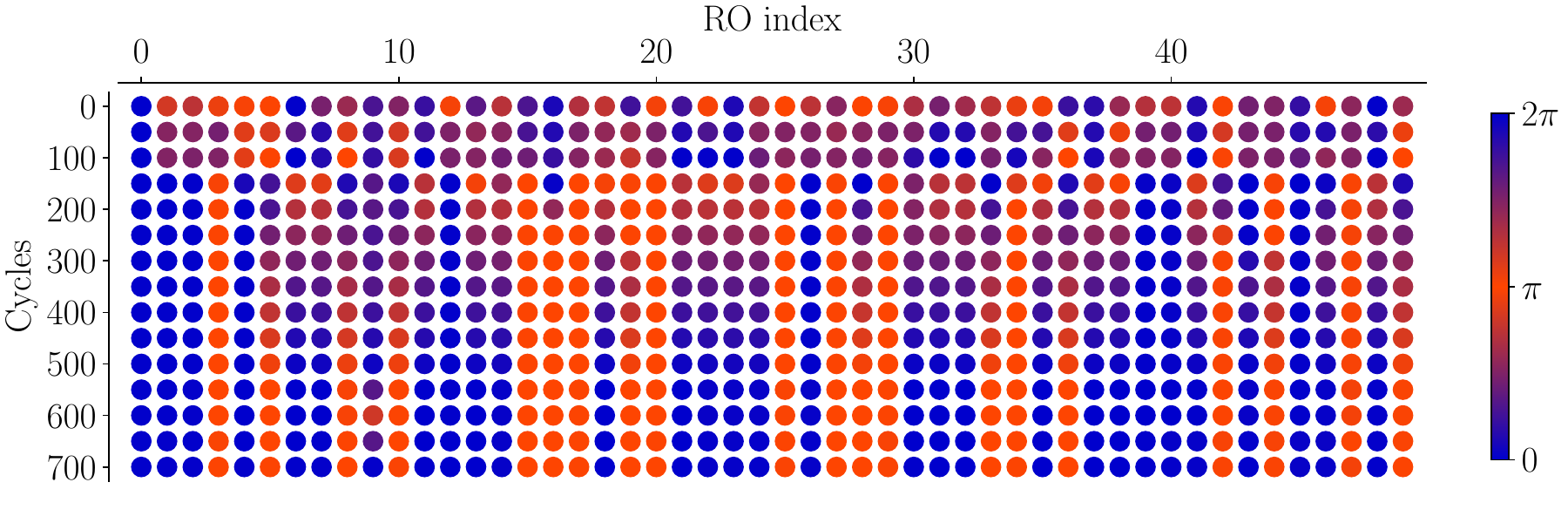}
    \vspace{-6mm}
    \caption{The evolution of RO-phases from random to binarized phases for a particular problem; phases are represented by the color bar at right.
    }
    \label{fig:phase_evolution}
    \vspace{-6mm}
\end{figure}

Fig.~\ref{fig:phase_evolution} shows the evolution of phases for a representative problem in a 50-spin A2A system, over multiple snapshots of time, starting from random phases. The blue and red colors correspond to the $\pm 1$ spins. As the state of the array evolves, the spins go through the continuum between the two states, shown by the color bar, until they converge to the final state, seen after 700 cycles.

\noindent
\underline{Comparison with hardware}
Since the A2A Ising hardware platform does not allow us to probe terminals to observe the states of the ROs, we compare the synchronized state from the hardware with that from a software simulation. The A2A hardware has a 50$\times$50 array~\cite{Lo2023}, using one RO from the array as a sampling clock and another as the reference spin. An arbitrary 48-spin Hamiltonian needs $(48+1)$ A2A ROs after accounting for the reference spin to implement the linear terms. Therefore, as a test set, we generate a set of 48-spin problems for five random values of graph density, \{0.2, 0.4, 0.6, 0.8, 1.0\}, with 50 problems for each density. As before, the coupling values for edges are chosen uniformly in $[-7, +7]-\{0\}$. For each problem, 100 samples obtained from the hardware are compared with DROID under randomized initial conditions.
We compare the distributions of Hamiltonian~\eqref{eq:ham}, for the sample of random coupling weights, for the solutions from the hardware solver and the simulator. 

\begin{figure}[htb]
\centering
\includegraphics[width=0.65\linewidth]{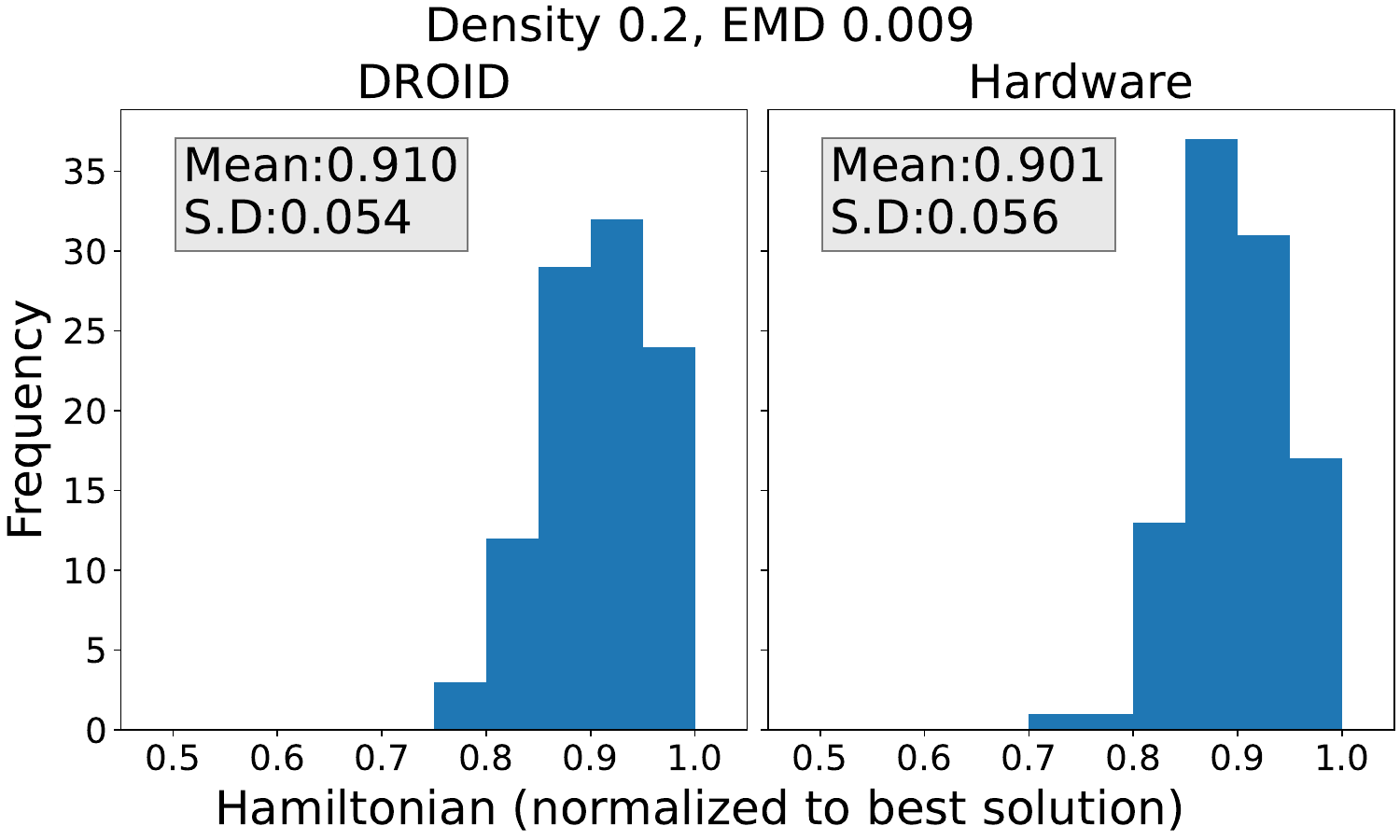}
\vspace{-4mm}
\caption{Distributions of the Hamiltonian from 100 samples of a problem of density 0.2 obtained from (a)~DROID and (b)~the hardware. The X-axis has been normalized to the best Hamiltonian value obtained for the problem.}
\label{fig:distribution}
\vspace{-4mm}
\end{figure}

The distributions of the optimized Hamiltonian from DROID and the hardware solver, for a specific problem instance with a density of 0.2, are shown in Fig.~\ref{fig:distribution}. The x-axis is normalized to the optimum Hamiltonian for the problem, and the bin size is 5\% of the optimum Hamiltonian value, i.e., 0.05 units on this normalized axis. Visually, the histograms are seen to be very similar.

Quantitatively, we use Earth Mover Distance (EMD)~\cite{Rubner98} to measure the similarity of the two distributions. The EMD quantifies the amount of work required to transform one distribution to the other and is robust to small changes to the bin size.  Intuitively, if we view one distribution as a mass of earth spread in space and the other as a collection of holes in the very same space, the EMD measures the least amount of work needed to fill the holes with earth; if the two distributions are identical, the work is zero~\cite{Rubner98}. A large EMD value suggests dissimilarity of distributions as more work is required to transform one distribution to the other, and therefore small values of EMD are desirable. The EMD between the distributions in Fig.~\ref{fig:distribution} is 0.009: this can be thought of as moving 9 of 100 samples over a distance of 10\% of the x-axis, which indicates that the distributions are close.

\input{tables/compare_distribution}

Table~\ref{tab:droid_hw_comparison} summarizes the EMD between the distributions across problems from the test set described earlier. The mean, minimum, maximum, and standard deviation of the EMD values for each graph density value correspond to a row in the table.  The small EMD values indicate that over the test set, the distribution of solutions from DROID resembles that from the hardware. 

%% file: tables/simulation_runtimes.tex
\begin{table}[H]
\vspace{-4mm}
\centering
\caption{Simulation runtimes for various array sizes.}
\label{tbl:runtime}
\resizebox{0.75\linewidth}{!}{
\begin{tabular}{|l|c|c|c|}
\hline
Array size & $5 \times 5$ & $20 \times 20$ & $50 \times 50 $ \\ \hline \hline
\# MOSFETs & 3100         & 52000          & 328000          \\ \hline
HSPICE     & 262s         & 1.04h          & 16.33h          \\ \hline
DROID      & 2.1s         & 3.5s           & 7.9s            \\ \hline
Speedup    & 125$\times$  & 1072$\times$   & 7441$\times$    \\ \hline
\end{tabular}
}
\end{table}

%% file: tables/compare_distribution.tex
\begin{table}[H]
    \centering
    \vspace{-3mm}
    \caption{EMD between distributions from DROID and A2A Ising chip for 5 graph densities, 50 problems/density, 100 samples/problem}
    \resizebox{0.7\linewidth}{!}{
    \begin{tabular}{|c|c|c|c|c|}
    \hline
    \multirow{2}{*}{Density} & \multicolumn{4}{c|}{EMD}                \\ \cline{2-5} 
                             & Mean    & Minimum  & Maximum   & S.D.   \\ \hline
    0.2                      & 0.021   & 0.005    & 0.073     & 0.016  \\ \hline
    0.4                      & 0.021   & 0.004    & 0.054     & 0.012  \\ \hline
    0.6                      & 0.014   & 0.002    & 0.036     & 0.009  \\ \hline
    0.8                      & 0.017   & 0.003    & 0.040     & 0.008  \\ \hline
    1.0                      & 0.016   & 0.001    & 0.044     & 0.009  \\ \hline \hline
    \multicolumn{1}{|c}{Average}
                             & \multicolumn{1}{c}{0.018}   
                                       & \multicolumn{1}{c}{}
                                                  & \multicolumn{1}{c}{} 
                                                              & 0.011  \\ \hline
    \end{tabular}
    }
    \label{tab:droid_hw_comparison}
\end{table}

%% file: sec/6_Conclusion.tex
\section{Conclusion}
\label{sec:Conclusion}

\noindent
This paper has presented DROID, an event-driven simulator for RO-based Ising machines. It is analytically shown that its formulation is more general than conventional continuous-time models, and that its event-driven nature captures effects that have not been addressed in previous analytical solvers. DROID is shown to be nearly four orders of magnitude faster, at similar accuracy on a $50 \times 50$ array, which has the computational power of a $\sim$2500-spin planar array. This speedup enables us to simulate the all-to-all RO array with multiple initial conditions and obtain distributions of solutions that simulate the hardware. 

%% file: sec/Appendix.tex
\section{}
\label{app:appendix}

\subsection{Pseudocode for the \textproc{process\_event} function}
\label{app:process_event}

\noindent
The pseudocode for the function \textproc{process\_event} is listed in Algorithm~\ref{alg:process_event}. The function receives an event E, the queue Q, and the maps Net2Event and PendingTrigger, and generates output event(s) which it inserts into Q. The major steps involved are as follows:

\noindent 
\textbf{Step PE1: Find the cell type of the cell that receives E as input} The netlist and the net name property of the event object E are used to determine the cell that receives the event at its input.  If the cell is an enable cell, the output event can be calculated directly (lines~\ref{alg:enable_start}--\ref{alg:enable_end}), and the event is processed. Otherwise, in case of a coupling or a shorting cell, we proceed to Step PE2. 

\noindent 
\textbf{Step PE2: Find the direction of the path that E lies on} If the event E occurs on a net on the reverse path (on pin $h^r_{in}$ or $v^r_{in}$ of a cell), there is no coupling interaction and the output events are calculated directly (lines~\ref{alg:backward_start}--\ref{alg:backward_end}). Otherwise, for an event on a forward path (on pin $h^f_{in}$ or $v^f_{in}$), we proceed to Step PE3.

\input{pseudocode/process_event}

\noindent 
\textbf{Step PE3: Find an interacting event for the forward path} If the event E occurs on $h_{in}^{f}$, then only an event at $v_{in}^{f}$ of the same cell can interact with E and vice-versa. The map Net2Event can be queried with the net name of the other input to look for an interacting event E'. Even if an event at the other input is not found in the map, it is possible that such an event has not yet been generated from a predecessor cell, and we proceed to Step PE4 to look for events that might still result in interaction with E. Note that the interaction window lies in a range of $\pm W$ around the arrival time of event E, and therefore this process should identify any event within this window -- one that precedes or succeeds E.   

If found, the switching information for events E and E' is sufficient to determine whether they interact or not, and the output events may be calculated (lines~\ref{alg:forward_start}--\ref{alg:forward_end}). Interacting events E and E' generate a pair of output events, while a noninteracting event generates one output event. 

\noindent
\textbf{Step PE4: Look back to find events that might result in interaction} When an event is on the other input not found in the map Net2Event, the function \textproc{look\_back}, described in detail in Appendix~\ref{app:appendix}-\ref{app:look_back}, is invoked.  As described at the end of the example, for an event E at a cell C, intuitively, this procedure looks into the predecessors of C to determine whether any incoming, but as yet unprocessed, event might result in another event that interacts with E. If such an event is found, the event E is added to the PendingTrigger map with the net name of the other input of the cell as its trigger. If not, we generate the output event from E assuming no interaction. 

\noindent
\textbf{Step PE5: Check if output events are triggers to a pending event} If E generates an event on a net that is a trigger to a pending event in the PendingTrigger map, then the pending event is ready to be processed and is added to the queue.

\subsection{Pseudocode for the \textproc{look\_back} function}
\label{app:look_back}

\input{pseudocode/look_back}
\noindent
The pseudocode for the function \textproc{look\_back} is provided in Algorithm~\ref{alg:look_back}. The inputs to the function are a net \textit{event\_net}, a window of arrival time (left, right), a threshold arrival time \textit{threshold}, and the map Net2Event. \textproc{look\_back} is a recursive function that terminates when it encounters one of three base cases:

\begin{itemize}[noitemsep,topsep=-1pt,leftmargin=*]
    \item The latest arrival time of an event at \textit{event\_net} is less than \textit{threshold}. The function returns false. 
    \item An event arrives at \textit{event\_net} within the window and can result in an interaction. The function returns true.
    \item An event arrives at \textit{event\_net} outside the window and will not result in an interaction. The function returns false.
\end{itemize}

The first base case prevents us from looking at too many predecessors by comparing the latest arrival time to \textit{threshold}. The \textit{threshold} is assigned a value in \textproc{process\_event} which corresponds to the earliest arrival time for an interaction with the event that invoked \textproc{look\_back}. 
The second and third cases require looking at the map Net2Event for an event with the key \textit{event\_net} to decide if it will arrive within the arrival time window. When none of the base cases are encountered, the function moves to the recursive step.

The predecessor of \textit{event\_net}, which we call \textit{preceding\_net} is found when an event at \textit{event\_net} does not exist in Net2Event. The cell that \textit{preceding\_net} is an input of is called \textit{preceding\_cell}. The minimum and maximum delays ($d_{min}$ and $d_{max}$, respectively) for \textit{preceding\_cell} are used to calculate a window of arrival for \textit{preceding\_net}. A recursive call is made at \textit{preceding\_net} with this new window, with the recursion concluding when the interaction window has been exceeded.

%% file: pseudocode/process_event.tex
\begin{algorithm}[H]
    {
    \small
    \caption{\textproc{process\_event}: Processing an event in the event queue}
    \label{alg:process_event}
    \algrenewcommand{\algorithmicindent}{0.5em}
        \begin{algorithmic}[1]
        \Function{process\_event}{E, Q, Net2Event, PendingTrigger}
            \State \textit{cell} =  the cell instance that receives event E at its input
            \State remove\_event = list() \textit{// List for processed events}
            \State NE = list() \textit{// List for output events}
            \State \textit{// Step PE1: Find the cell type of the cell that receives E as input.} 
            \If {\textit{cell} is a coupling or shorting cell}
                \State \textit{pin} = the pin of \textit{cell} that event E occurs at
                \State \textit{// Step PE2: Find the direction of the path that E lies on.}
                \If {\textit{pin} is $h_{in}^{f}$ or $v_{in}^{f}$}
                    \State \textit{// Step PE3: Find an interacting event for the forward path.}
                    \If {event E' exists on other input} \label{alg:forward_start}
                        \If {E and E' are within a window of W}
                            \State NE $\xleftarrow{}$ Calculate output events from E and E'
                            \State remove\_event.add([E, E'])
                        \Else
                            \State NE $\xleftarrow{}$ Calculate output event from E
                            \State remove\_event.add(E)
                        \EndIf \label{alg:forward_end}
                    \Else
                        \State \textit{// Step PE4: Look back to find events that might result in interaction.}
                        \State \textit{net} = the net connected to other input of cell
                        \State (left, right) = (E.arrival\_time - W, E.arrival\_time + W)
                        \State status = \Call{look\_back}{\textit{net}, (left, right), left, Net2Event}
                        \If {status}
                            \State PendingTrigger[\textit{net}] = E
                        \Else
                            \State NE $\xleftarrow{}$ Calculate output event from E
                            \State remove\_event.add(E)
                        \EndIf
                    \EndIf
                \Else
                    \State \textit{// Calculate events on the backward path.} \label{alg:backward_start}
                    \State NE $\xleftarrow{}$ Calculate output event from E
                    \State remove\_event.add(E) \label{alg:backward_end}
                \EndIf
            \Else 
                \State \textit{// enable cell} \label{alg:enable_start}
                \State NE $\xleftarrow{}$ Calculate output event from E
                \State remove\_event.add(E) \label{alg:enable_end}
            \EndIf
            \State \textit{// Step PE5: Check if events in NE are triggers to a pending event}
            \If {NE not empty} 
                \For {new\_event $\in$ NE}
                    \State Q.add(new\_event)
                    \State Net2Event[new\_event.netname] = new\_event
                    \If {new\_event.netname \textbf{in} PendingTrigger}
                        \State Q.add(PendingTrigger[new\_event.netname])
                        \State PendingTrigger.pop(new\_event.netname)
                    \EndIf
                \EndFor
            \EndIf
            \State \textit{// Remove consumed events}
            \For {used\_event $\in$ remove\_events}
                \State Net2Event.pop(used\_event.netname)
            \EndFor
        \EndFunction
        \end{algorithmic}
    }
\end{algorithm}

%% file: pseudocode/look_back.tex
\begin{algorithm}[H]
    {
    \small
    \caption{\textproc{look\_back}: Identification of events that might interact with an event being processed}
    \label{alg:look_back}
    \algrenewcommand{\algorithmicindent}{0.5em} 
    \begin{algorithmic}[1]
    \Function{look\_back}{\textit{event\_net}, (left, right), \textit{threshold}, Net2Event}
        \State \textit{// looks back recursively to find events that might interact}
        \If {right $<$ \textit{threshold}} 
            \State \textit{// Base case: looked back far enough}
            \State \Return False 
        \EndIf
        \If {\textit{event\_net} is in Net2Event}
            \State \textit{// Base cases: Interacting or not}
            \If {the event will arrive in (left, right)}
                \State \Return True 
            \ElsIf {the event will not arrive in (left, right)}
                \State \Return False 
            \EndIf
        \Else
            \State \textit{// Recursive step: need to look back further}
            \State \textit{preceding\_net} = the net that can cause an event at \textit{event\_net}
            \State \textit{preceding\_cell} = the instance for which event\_net is an output
            \State (nleft, nright) = (left - \textit{preceding\_cell}.$d_{max}$, right - \textit{preceding\_cell}.$d_{min}$)
            \State \Return \Call{look\_back}{\textit{preceding\_net}, (nleft, nright), \textit{threshold}, Net2Event}
        \EndIf
    \EndFunction
    \end{algorithmic}
    }
\end{algorithm}

%% file: main.bbl
\begin{thebibliography}{10}
\providecommand{\url}[1]{#1}
\csname url@samestyle\endcsname
\providecommand{\newblock}{\relax}
\providecommand{\bibinfo}[2]{#2}
\providecommand{\BIBentrySTDinterwordspacing}{\spaceskip=0pt\relax}
\providecommand{\BIBentryALTinterwordstretchfactor}{4}
\providecommand{\BIBentryALTinterwordspacing}{\spaceskip=\fontdimen2\font plus
\BIBentryALTinterwordstretchfactor\fontdimen3\font minus \fontdimen4\font\relax}
\providecommand{\BIBforeignlanguage}[2]{{%
\expandafter\ifx\csname l@#1\endcsname\relax
\typeout{** WARNING: IEEEtran.bst: No hyphenation pattern has been}%
\typeout{** loaded for the language `#1'. Using the pattern for}%
\typeout{** the default language instead.}%
\else
\language=\csname l@#1\endcsname
\fi
#2}}
\providecommand{\BIBdecl}{\relax}
\BIBdecl

\bibitem{Lucas_Ising_Frontiers14}
A.~Lucas, ``{Ising formulations of many {NP} problems},'' \emph{Frontiers in Physics}, vol.~2, pp. 5:1--5:15, Feb. 2014.

\bibitem{Johnson2011}
M.~W. Johnson \emph{et~al.}, ``{Quantum annealing with manufactured spins},'' \emph{Nature}, vol. 473, no. 7346, pp. 194--198, May 2011.

\bibitem{Bian2014}
Z.~Bian \emph{et~al.}, ``{Discrete optimization using quantum annealing on sparse {Ising} models},'' \emph{Frontiers in Physics}, vol.~2, pp. {56:1--56:10}, Sep 2014.

\bibitem{Inagaki2016}
T.~Inagaki \emph{et~al.}, ``{A coherent {Ising} machine for 2000-node optimization problems},'' \emph{Science}, vol. 354, no. 6312, pp. 603--606, 2016.

\bibitem{Yamamoto2017}
Y.~Yamamoto \emph{et~al.}, ``{Coherent {Ising} machines---optical neural networks operating at the quantum limit},'' \emph{npj Quantum Information}, vol.~3, no.~1, pp. 49:1--49:15, Dec 2017.

\bibitem{wang2019matlab}
T.~Wang \emph{et~al.}, ``{OIM: Oscillator-based Ising Machines for Solving Combinatorial Optimisation Problems},'' 2019, \url{https://arxiv.org/abs/1903.07163}.

\bibitem{moy20221}
W.~Moy \emph{et~al.}, ``A 1,968-node coupled ring oscillator circuit for combinatorial optimization problem solving,'' \emph{Nature Electronics}, vol.~5, no.~5, pp. 310--317, May 2022.

\bibitem{Lo2023}
H.~Lo \emph{et~al.}, ``{An Ising solver chip based on coupled ring oscillators with a 48-node all-to-all connected array architecture},'' \emph{Nature Electronics}, vol.~6, no.~10, pp. 771--778, Oct 2023.

\bibitem{Yamaoka16}
M.~Yamaoka \emph{et~al.}, ``A {20K}-spin {Ising} chip to solve combinatorial optimization problems with {CMOS} annealing,'' \emph{IEEE Journal of Solid-State Circuits}, vol.~51, no.~1, pp. 303--309, Jan. 2016.

\bibitem{Willms17}
A.~R. Willms \emph{et~al.}, ``Huygens' clocks revisited,'' \emph{Royal Society Open Science}, vol.~4, pp. 170\,777:1--170\,777:33, 2017.

\bibitem{Adler}
R.~Adler, ``{A Study of Locking Phenomena in Oscillators},'' \emph{Proceedings of the IRE}, vol.~34, no.~6, pp. 351--357, 1946.

\bibitem{WINFREE196715}
A.~T. Winfree, ``{Biological rhythms and the behavior of populations of coupled oscillators},'' \emph{Journal of Theoretical Biology}, vol.~16, no.~1, pp. 15--42, 1967.

\bibitem{Kuramoto1984-hj}
Y.~Kuramoto, \emph{Chemical Oscillations, Waves, and Turbulence}.\hskip 1em plus 0.5em minus 0.4em\relax Berlin, Germany: Springer, 1984.

\bibitem{Bhansali2009}
P.~Bhansali \emph{et~al.}, ``{Gen-{A}dler: The generalized {A}dler's equation for injection locking analysis in oscillators},'' in \emph{Proceedings of the Asia-South Pacific Design Automation Conference}, 2009, pp. 522--527.

\bibitem{sreedhara23}
S.~Sreedhara \emph{et~al.}, ``{MU-MIMO Detection Using Oscillator Ising Machines},'' in \emph{Proceedings of the IEEE/ACM International Conference on Computer-Aided Design}, 2023.

\bibitem{sreedhara_date23}
------, ``{Digital Emulation of Oscillator Ising Machines},'' in \emph{Proceedings of the Design, Automation \& Test in Europe}, 2023.

\bibitem{cilasun2024}
H.~Cılasun \emph{et~al.}, ``{COBI: A Coupled Oscillator Based Ising Chip for Combinatorial Optimization},'' \emph{{ResearchSquare}}, 2024.

\bibitem{Sapatnekar04}
S.~Sapatnekar, \emph{Timing}.\hskip 1em plus 0.5em minus 0.4em\relax New York, NY: Springer, 2004.

\bibitem{Ahmed2021}
I.~Ahmed \emph{et~al.}, ``{A Probabilistic Compute Fabric Based on Coupled Ring Oscillators for Solving Combinatorial Optimization Problems},'' \emph{IEEE Journal of Solid-State Circuits}, vol.~56, no.~9, pp. 2870--2880, 2021.

\bibitem{Tabi21}
Z.~I. Tabi \emph{et~al.}, ``Evaluation of quantum annealer performance via the massive {MIMO} problem,'' \emph{IEEE Access}, vol.~9, pp. 131\,658--131\,671, 2021.

\bibitem{Lucas2019}
A.~Lucas, ``Hard combinatorial problems and minor embeddings on lattice graphs,'' \emph{Quantum Information Processing}, vol.~18, no.~7, pp. 203:1--203:38, May 2019.

\bibitem{cilasun20243sat}
H.~C{\i}lasun \emph{et~al.}, ``{3SAT} on an all-to-all-connected {CMOS} {Ising} solver chip,'' \emph{Scientific Reports}, vol.~14, no.~1, pp. 10\,757:1--10\,757:11, 2024.

\bibitem{Rubner98}
Y.~Rubner \emph{et~al.}, ``A metric for distributions with applications to image databases,'' in \emph{Proceedings of the IEEE International Conference on Computer Vision}, 1998, pp. 59--66.

\end{thebibliography}
